\documentclass{article}

\usepackage{arxiv}

\usepackage[utf8]{inputenc} 
\usepackage[T1]{fontenc}    
\usepackage{lmodern}        
\usepackage{natbib}
\setcitestyle{authoryear, open={(}, close={)}}
\usepackage{hyperref}       
\usepackage{url}            
\usepackage{booktabs}       
\usepackage{amsfonts}       
\usepackage{nicefrac}       
\usepackage{microtype}      
\usepackage{graphicx}

\title{Projection predictive variable selection for discrete response families with finite support}

\author{
    Frank Weber
   \\
    Institute for Biostatistics and Informatics in Medicine and Ageing Research \\
    Rostock University Medical Center \\
  Rostock, Germany \\
  \texttt{\href{mailto:frank.weber@uni-rostock.de}{\nolinkurl{frank.weber@uni-rostock.de}}} \\
   \And
    Änne Glass
   \\
    Institute for Biostatistics and Informatics in Medicine and Ageing Research \\
    Rostock University Medical Center \\
  Rostock, Germany \\
  \texttt{\href{mailto:aenne.glass@uni-rostock.de}{\nolinkurl{aenne.glass@uni-rostock.de}}} \\
   \And
    Aki Vehtari
   \\
    Department of Computer Science \\
    Aalto University \\
  Espoo, Finland \\
  \texttt{\href{mailto:aki.vehtari@aalto.fi}{\nolinkurl{aki.vehtari@aalto.fi}}} \\
  }

\usepackage{microtype}
\usepackage{amsmath}
\usepackage{tikz}
\usepackage{xcolor}
\hypersetup{colorlinks, linkcolor={red!50!black}, citecolor={red!50!black}, urlcolor={blue!80!black}}
\usepackage{adjustbox}
\usepackage[title]{appendix}
\usepackage{chngcntr}

\begin{document}
\maketitle

\newcommand{\appref}[1]{Appendix~#1}
\newcommand{\maxdiff}{\appref{\ref{maxdiff}}}
\newcommand{\resabs}{\appref{\ref{resabs}}}
\newcommand{\pkg}{\textbf}
\newcommand{\mainref}[1]{Figure~#1}
\newcommand{\figauglat}{\mainref{\ref{fig:aug-lat}}}
\newcommand{\figdiff}{\mainref{\ref{fig:diff}}}
\newcommand{\figdiffse}{\mainref{\ref{fig:diff-se}}}
\newcommand{\figdiffsgg}{\mainref{\ref{fig:diff-sgg}}}

\begin{abstract}
The projection predictive variable selection is a decision-theoretically justified Bayesian variable selection approach achieving an outstanding trade-off between predictive performance and sparsity. Its projection problem is not easy to solve in general because it is based on the Kullback-Leibler divergence from a restricted posterior predictive distribution of the so-called reference model to the parameter-conditional predictive distribution of a candidate model. Previous work showed how this projection problem can be solved for response families employed in generalized linear models and how an approximate latent-space approach can be used for many other response families. Here, we present an exact projection method for all response families with discrete and finite support, called the augmented-data projection. A simulation study for an ordinal response family shows that the proposed method performs better than or similarly to the previously proposed approximate latent-space projection. The cost of the slightly better performance of the augmented-data projection is a substantial increase in runtime. Thus, in such cases, we recommend the latent projection in the early phase of a model-building workflow and the augmented-data projection for final results. The ordinal response family from our simulation study is supported by both projection methods, but we also include a real-world cancer subtyping example with a nominal response family, a case that is not supported by the latent projection.
\end{abstract}


\section{Introduction}\label{intro}

The projection predictive variable selection \citep{piironen_projective_2020, catalina_projection_2022} is a special predictive model selection method \citep{vehtari_survey_2012} for Bayesian regression models that comes with valid post-selection inference (disregarding the selection of the final model size) and has been shown to perform better---in general---than alternative methods \citep{piironen_comparison_2017}.
It is based on the Bayesian decision-theoretical variable selection framework by \citet{lindley_choice_1968} and the practical draw-by-draw Kullback-Leibler (KL) projection proposed by \citet{goutis_model_1998} and \citet{dupuis_variable_2003}.
So far, the implementation of the projection predictive variable selection in the R \citep{r_core_team_r_2022} package \pkg{projpred}\footnote{Currently, \pkg{projpred} may be regarded as the most popular Bayesian variable selection package for R. This can be checked by comparing the download numbers for \pkg{projpred}, \pkg{BayesVarSel} \citep{garcia-donato_bayesian_2018}, \pkg{BAS} \citep{clyde_bas_2022}, \pkg{varbvs} \citep{carbonetto_scalable_2012}, \pkg{spikeSlabGAM} \citep{scheipl_spikeslabgam_2011}, \pkg{BVSNLP} \citep{nikooienejad_bvsnlp_2020}, \pkg{ptycho} \citep{stell_ptycho_2015}, \pkg{BayesSUR} \citep{zhao_bayessur_2021}, \pkg{BGLR} \citep{perez_genome-wide_2014}, \pkg{MBSGS} \citep{liquet_mbsgs_2017}, and \pkg{mombf} \citep{rossell_mombf_2023} via \pkg{cranlogs} \citep{csardi_cranlogs_2019}. Last check: April 26, 2023.} \citep{piironen_projpred_2022} has been restricted to the Gaussian, the binomial, and the Poisson response families.
Recently, the \emph{latent projection} \citep{catalina_latent_2021} has extended the range of possible response families considerably, for example, to the ordinal family underlying \texttt{MASS::polr()} \citep{venables_modern_2002}.
However, the latent projection is an approximate approach as it replaces the original projection problem with a latent projection problem.
Here (section~\ref{math}), we present the exact solution to the original projection problem for discrete finite-support response families and call the corresponding procedure the \emph{augmented-data projection}.

For investigating the performance of the augmented-data projection (section~\ref{sim}), we confine ourselves to a simulation study comparing the augmented-data projection to the latent projection because the generally superior performance of the projection predictive variable selection based on the traditional projection and based on the latent projection has already been demonstrated by \citet{piironen_comparison_2017} and \citet{catalina_latent_2021}, respectively.

We illustrate the application of the augmented-data projection in section~\ref{ex} by the help of a real-world example, thereby also demonstrating another benefit of the augmented-data projection, namely the support for some response families which are not supported by the latent projection.

Finally, our work is discussed in section~\ref{disc}, where we also mention possible modifications of the augmented-data projection to extend it to more response families in the future.

\section{Augmented-data projection}\label{math}

\subsection{Notation}\label{math-notation}

For the following mathematical presentation of the augmented-data projection (a special case of the general approach that is presented first), we assume the availability of a dataset with \(N\) observations.
The observed response vector will be denoted by \(\boldsymbol{y} = (y_1, \dotsc, y_N)^{\scriptscriptstyle \mathsf{T}} \in \mathcal{Y}^{N} \subseteq \mathbb{R}^{N}\).
We do not introduce any notation for the corresponding predictor data as we will always be conditioning implicitly on it.
By \(\boldsymbol{\tilde{y}} = (\tilde{y}_1, \dotsc, \tilde{y}_N)^{\scriptscriptstyle \mathsf{T}}\), we will denote \emph{unobserved} response values at the same observed predictor values, with realizations in \(\mathcal{Y}^{N}\).

A crucial part \citep{piironen_projective_2020, pavone_using_2022} for the superior performance of the projection predictive variable selection is the \emph{reference model}, which is the best possible model (in terms of predictive performance) one can construct.
For \pkg{projpred}, the reference model is usually fitted within \pkg{rstanarm} \citep{goodrich_rstanarm_2022} or \pkg{brms} \citep{burkner_brms_2017, burkner_advanced_2018} which both rely on Stan \citep{carpenter_stan_2017, stan_development_team_stan_2022}, a probabilistic programming language and software that is mainly used for its dynamic Hamiltonian Monte Carlo (HMC) algorithm, a modern Markov chain Monte Carlo (MCMC) sampler.
However, the methodology behind \pkg{projpred} is more general and does not require the reference model to be fitted within \pkg{rstanarm} or \pkg{brms}.
Thus, we start by assuming to have \(S^{*}\) draws \(\boldsymbol{\theta}^{*}_{s} \in \boldsymbol{\Theta}^{*}\) (\(s \in \{1, \dotsc, S^{*}\}\)) from the reference model's posterior distribution, with \(\boldsymbol{\Theta}^{*}\) denoting the reference model's parameter space.
Furthermore, we assume that these \(S^{*}\) posterior draws have been clustered or thinned so that \(\{1, \dotsc, S^{*}\} \supseteq \mathop{\bigcup}_{c = 1}^{C} \mathcal{I}^{*}_{c}\) with disjoint index sets \(\mathcal{I}^{*}_{c}\).
An explanation how the clustering is performed in \pkg{projpred} will be given below.
Based on the clustering (or thinning), we can define the reference model's \(c\)-restricted posterior predictive distribution (for observation \(i\)):
\[p(\tilde{y}_i|\mathcal{I}^{*}_{c}) = \frac{1}{|\mathcal{I}^{*}_{c}|} \sum_{s \in \mathcal{I}^{*}_{c}} p(\tilde{y}_i|\boldsymbol{\theta}^{*}_{s}).\]
In doing so, the conditioning on an index set is slightly abusing notation, but we think it improves readability while at the same time reflecting the basic idea behind this empirical average.
Expectations with respect to \(p(\tilde{y}_i|\mathcal{I}^{*}_{c})\) will be denoted by \(\mathbb{E}(\cdot|\mathcal{I}^{*}_{c})\).

A model selection problem comes with several \emph{candidate models}, of which we will consider only a single one here, to avoid cluttering notation.
In the context of a variable selection problem, this candidate model may also be called a \emph{submodel} of the full model which includes all predictors.
The parameter space of this representative submodel will be denoted by \(\boldsymbol{\Theta}\) and its parameter-conditional predictive distribution (i.e., its likelihood when regarded as a function of the parameters) by \(p(\tilde{y}_i|\boldsymbol{\theta})\) (for \(\boldsymbol{\theta} \in \boldsymbol{\Theta}\)).
We emphasize that in general, \(\boldsymbol{\Theta}\) does not have to be related to \(\boldsymbol{\Theta}^{*}\) in any form (in particular, it does not have to be a restricted subspace).

Finally, we need the Kullback-Leibler (KL) divergence \citep{kullback_information_1951} from a distribution \(p(x)\) to a distribution \(q(x)\):
\[D_{\mathrm{KL}}\!\left(p(x) \:\middle\|\: q(x) \right) = \mathbb{E}_{p(x)}\!\left(\log \frac{p(x)}{q(x)}\right)\]
where we have added the subscript \(p(x)\) to clarify the distribution that the expectation refers to.

For the clustering (and several other steps), \pkg{projpred} requires an invertible link function \(g\).
With this link function \(g\), \pkg{projpred} performs the clustering of the \(S^{*}\) posterior draws by applying \texttt{stats::kmeans()} (the \pkg{stats} package is part of R) to the \(S^{*}\) length-\(N\) vectors \(\boldsymbol{g}(\mathbb{E}(\boldsymbol{\tilde{y}}|\boldsymbol{\theta}^{*}_{s}))\) where \(\boldsymbol{g}\) denotes the vectorized link function, i.e., the function which applies \(g\) to each element of a vector.

\subsection{General approach}\label{math-general}

In general, the submodel's projected parameter values for cluster (or thinned draw) \(c \in \{1, \dotsc, C\}\) are obtained by solving
\begin{align}
  \boldsymbol{\theta}_{c} = \;&\mathop{\mathrm{argmin}}\limits_{\boldsymbol{\theta} \in \boldsymbol{\Theta}}
  \frac{1}{N} \sum_{i = 1}^{N}
  D_{\mathrm{KL}}\!\left(
  p(\tilde{y}_i|\mathcal{I}^{*}_{c})
  \:\middle\|\:
  p(\tilde{y}_i|\boldsymbol{\theta})
  \right) \notag \\
  = \;&\mathop{\mathrm{argmax}}\limits_{\boldsymbol{\theta} \in \boldsymbol{\Theta}}
  \sum_{i = 1}^{N}
  \mathbb{E}\!\left(\log p(\tilde{y}_i|\boldsymbol{\theta}) \:\middle|\: \mathcal{I}^{*}_{c}\right), \label{eqn:proj-general}
\end{align}
see \citet{piironen_projective_2020}.

This projection problem is not easy to solve in general because \(\mathbb{E}(\cdot|\mathcal{I}^{*}_{c})\) is an expectation with respect to \(p(\tilde{y}_i|\mathcal{I}^{*}_{c})\).
Equation~(\ref{eqn:proj-general}) simplifies a lot if the submodel's response family follows the definition from \citet[equation (2.4)]{mccullagh_generalized_1989} because in that case, \(\log p(\tilde{y}_i|\boldsymbol{\theta})\) is linear in \(\tilde{y}_i\), at least for optimization with respect to the non-dispersion parameters.
Another simplifying case is \(|\mathcal{Y}| < \infty\), which is the gist here (see section~\ref{math-finite}).

\subsection{Discrete finite-support response families}\label{math-finite}

In case of \(|\mathcal{Y}| < \infty\), equation~(\ref{eqn:proj-general}) simplifies because \(\mathbb{E}(\cdot|\mathcal{I}^{*}_{c})\) is then a sum over all possible response values:
\begin{align}
  \boldsymbol{\theta}_{c} = \;&\mathop{\mathrm{argmax}}\limits_{\boldsymbol{\theta} \in \boldsymbol{\Theta}}
  \sum_{i = 1}^{N}
  \sum_{\tilde{y} \in \mathcal{Y}}
  a^{*}_{c, i, \tilde{y}}
  \log p(\tilde{y}_i = \tilde{y}|\boldsymbol{\theta}) \label{eqn:proj-aug}
\end{align}
with \(a^{*}_{c, i, \tilde{y}} = p(\tilde{y}_i = \tilde{y}|\mathcal{I}^{*}_{c})\).
Equation~(\ref{eqn:proj-aug}) is simply a weighted maximum-likelihood (ML) problem when using an augmented dataset where each observation is repeated \(|\mathcal{Y}|\) times and the response value is set to each possible value \(\tilde{y} \in \mathcal{Y}\) in turn so that the resulting augmented dataset has a total of \(N \cdot |\mathcal{Y}|\) rows.
This approach is what we call the augmented-data projection, although for implementation in \pkg{projpred}, the augmented dataset is constructed internally to have \(|\mathcal{Y}|\) blocks of \(N\) rows instead of the other way round.

Equation~(\ref{eqn:proj-aug}) shows that the augmented-data projection consists of fitting to the fit of the reference model, a fundamental property already exhibited by the traditional projection \citep{piironen_projective_2020}.
In case of a discrete response family with finite support, the fit of the reference model just needs to be expressed differently, namely in terms of probabilities for all of the response categories, and fitting to that fit then needs to be done in a \emph{weighted} fashion.

Due to the augmented-data projection being a weighted ML problem, the basic idea for implementing it in \pkg{projpred} is simply to apply existing R functions capable of performing a weighted ML estimation (e.g., \texttt{MASS::polr()} in case of the commonly used cumulative ordinal models) to the augmented dataset.
Currently, \pkg{projpred}'s augmented-data projection adds support for the \texttt{brms::cumulative()} family, for \texttt{rstanarm::stan\_polr()} fits, and for the \texttt{brms::categorical()} family.
(These families are additional in comparison to \pkg{projpred}'s traditional projection; \pkg{projpred}'s latent projection already supports these families, except for the \texttt{brms::categorical()} family.)
We emphasize that these families refer to the \emph{submodels}, not the reference model.
Typically, the reference model has the same response family as the submodels.
In general, the reference model is allowed to have a different family.
In case of the augmented-data projection, the only requirement concerning the form of the reference model is that its response family is discrete and has finite support (otherwise, the step from equation~(\ref{eqn:proj-general}) to equation~(\ref{eqn:proj-aug}) would be incorrect).
(In theory, equation~(\ref{eqn:proj-aug}) does not require the \emph{submodel} to have a discrete finite-support response family, but typically---and especially with respect to the implementation in \pkg{projpred}---this requirement makes sense.)

The augmented-data projection has been added in version 2.4.0 of \pkg{projpred} \citep{piironen_projpred_2022}.
In that version, an updated implementation of the latent projection \citep[compared to][]{catalina_latent_2021} has been included as well.
Note that for applying both---the augmented-data projection and the updated implementation of the latent projection---to reference model fits from \pkg{brms}, version 2.19.0 (or later) of \pkg{brms} is needed.

\section{Simulation study}\label{sim}

For the following simulation study comparing augmented-data and latent projection, we assume that the reader is familiar with the typical \pkg{projpred} workflow, as presented in the main vignette of the \pkg{projpred} package, for example.

\subsection{Setup}\label{sim-setup}

Since the latent projection does not support the \texttt{brms::categorical()} family, our simulation study is restricted to the \texttt{brms::cumulative()} family (which encodes the same observation model as in \texttt{rstanarm::stan\_polr()} fits).

More specifically, to comply with \citet{catalina_latent_2021}, we use \(J = |\mathcal{Y}| = 5\) response categories and the probit link function \(g = \Phi^{-1}\) (the quantile function of the standard normal distribution).
The number of observations is set to \(N = 100\), in accordance with the value used throughout the main article of \citet{catalina_latent_2021}.

Then, for each of \(R = 100\) simulation iterations, the simulation study involves the following steps:

\begin{enumerate}
\item
  Define the \(J - 1\) latent thresholds (intercepts) \(\zeta_{j}\) (\(j \in \{1, \dotsc, J - 1\}\)) as
  \[
   \zeta_{j} = g\!\left(\frac{j}{J}\right).
   \]
\item
  Generate \(P = 50\) regression coefficients \(\beta_{p}\) (\(p \in \{1, \dotsc, P\}\)) according to a regularized horseshoe prior \citep{piironen_sparsity_2017}.
  The underlying mechanism may be found in the R code for this simulation study (see the link at the end of this section).
  Here, we choose a global scale parameter of
  \[
   \tau_{0} = \frac{p_{0}}{P - p_{0}} \cdot \frac{\tilde{\sigma}}{\sqrt{N}}
   \]
  with \(p_{0} = 10\) and
  \[
   \tilde{\sigma}^2
   = \exp\!\left(\frac{1}{J} \sum_{j = 1}^{J} \log \tilde{\sigma}_{j}^2\right)
   = \sqrt[J]{\prod_{j = 1}^{J} \tilde{\sigma}_{j}^2},
   \]
  where \(\tilde{\sigma}_{j}^2\) are calculated according to Section 3.5 of \citet{piironen_sparsity_2017}, taking the same thresholds \(\zeta_{j}\) as defined above and assuming a typical data point with a latent predictor of zero so that all response categories are equally likely \citep[in analogy to the approach of][in case of the binomial family with the logit link]{piironen_hyperprior_2017}.
  Here, we obtain an overall pseudo variance of \(\tilde{\sigma}^2 \approx 1.06^2\).
  For the Student-\(t\) slab of the regularized horseshoe prior, we choose \(100\) degrees of freedom (effectively yielding a Gaussian slab) and a scale parameter of \(1\).
\item
  Generate a training dataset according to the following data-generating model where \(i \in \{1, \dots, N\}\):
  \begin{gather*}
     x_{i, p} \sim \mathcal{N}(0, 1) \quad (p \in \{1, \dots, P\}),\\
     \eta_i = \sum_{p = 1}^{P} \beta_{p} x_{i, p},\\
     \boldsymbol{\zeta} = (\zeta_{1}, \dotsc, \zeta_{J - 1})^{\scriptscriptstyle \mathsf{T}},\\
     y_i \sim \mathrm{Cumul}(\boldsymbol{\zeta}, \eta_i),
   \end{gather*}
  where \(\mathcal{N}(\mu, \sigma)\) denotes a normal distribution with mean \(\mu\) and standard deviation \(\sigma\) and \(\mathrm{Cumul}(\boldsymbol{\zeta}, \eta_i)\) denotes the distribution with probability mass function
  \begin{align*}
     p(y_i = j|\boldsymbol{\zeta}, \eta_i) = g^{-1}(\zeta_{j} - \eta_i) - g^{-1}(\zeta_{j - 1} - \eta_i)
   \end{align*}
  for \(j \in \{1, \dots, J\} = \mathcal{Y}\), exploiting auxiliary elements \(\zeta_{0} = -\infty\) and \(\zeta_{J} = \infty\) and defining \(g^{-1}(-\infty) = 0\) and \(g^{-1}(\infty) = 1\) (as well as \(\pm\infty - b = \pm\infty\) for \(b \in \mathbb{R}\)).
\item
  Generate an independent test dataset using the same data-generating model and the same settings (in particular, the same number \(N\) of observations) as for the training data.
\item
  Fit a reference model to the training data, using the data-generating model as the data-fitting model, except that the prior for the thresholds \(\zeta_{j}\) (\(j \in \{1, \dotsc, J - 1\}\)) is set to \(\mathcal{N}(0, 2.5)\) in the data-fitting model.
  The reference model fit is performed by \texttt{brms::brm()}, using the \pkg{cmdstanr} \citep{gabry_cmdstanr_2022} backend.
  We use the default of \(4\) Markov chains, each running \(1000\) warmup and \(1000\) post-warmup iterations.
  In order to avoid spurious divergences of Stan's dynamic HMC sampler, we aim at smaller step sizes by setting \texttt{adapt\_delta~=~0.99}.
  By specifying \texttt{init~=~1}, we narrow down the range that the initial parameter values are randomly drawn from (this was necessary to avoid that occasionally, some chains would initialize in an area of the parameter space with log posterior density numerically equal to \(-\infty\) or---shortly after initialization---would run into such an area).
  We checked the convergence of the Markov chains for an initial reference model fit (based on a dataset independent of those from the \(R = 100\) simulation iterations) by the help of common MCMC diagnostics \citep{betancourt_conceptual_2018, vehtari_rank-normalization_2021, stan_development_team_runtime_2022, burkner_posterior_2022}.
\item
  Run \pkg{projpred}.
  More specifically, the following steps are performed twice (once with the augmented-data projection and once with the latent projection, but based on the same training and test data and based on the same reference model fit):

  \begin{enumerate}
  \item
    Run \texttt{projpred::varsel()}, specifying the test data via argument \texttt{d\_test}.
    As search method, we choose the forward search because \pkg{projpred}'s augmented-data projection currently does not support the L1 search and also because the L1 search is often less accurate.
    Apart from that, we leave all other arguments at their default.
  \item
    For each submodel size along the solution path: Retrieve the mean log predictive density~(MLPD; actually mean log predictive probability, but the same acronym is used for simplicity), \(\Delta\mathrm{MLPD} = \mathrm{MLPD} - \mathrm{MLPD}^{*}\) (with \(\mathrm{MLPD}^{*}\) denoting the reference model MLPD), and the corresponding standard errors~(SEs).
    This is achieved via \texttt{projpred:::summary.vsel()}, once with \texttt{deltas~=~FALSE} (for the MLPD) and once with \texttt{deltas~=~TRUE} (for \(\Delta\mathrm{MLPD}\)).
    Here, the MLPD is the chosen performance statistic because of the desirable properties of the log score in general \citep{vehtari_survey_2012} and because \(\exp(\mathrm{MLPD})\), the geometric mean predictive density (GMPD), has an interpretable scale of \((0, 1)\) in case of a discrete response family.
    We denote MLPD based on the augmented-data projection by \(\mathrm{MLPD}_{\mathrm{aug}}\) and the corresponding \(\Delta\mathrm{MLPD}\) value by \(\Delta\mathrm{MLPD}_{\mathrm{aug}}\).
    For the latent projection, these are denoted by \(\mathrm{MLPD}_{\mathrm{lat}}\) and \(\Delta\mathrm{MLPD}_{\mathrm{lat}}\), respectively.
  \item
    Suggest a submodel size via \texttt{projpred::suggest\_size()}.
    As underlying performance statistic, we choose the MLPD again, for consistency with the results retrieved from \texttt{projpred:::summary.vsel()}.
    We denote the suggested size based on the augmented-data projection by \(G_{\mathrm{aug}}\) and the suggested size based on the latent projection by \(G_{\mathrm{lat}}\).
  \end{enumerate}
\end{enumerate}

The R code for this simulation study is available on \href{https://github.com/fweber144/simauglat/tree/fab1d3e2e2}{GitHub}\footnote{\url{https://github.com/fweber144/simauglat/tree/fab1d3e2e2}}. Figures were created with \pkg{ggplot2} \citep{wickham_ggplot2_2016}.

\subsection{Results}\label{sim-res}

\begin{figure}[!tb]
  \centering
  \begin{adjustbox}{max width=\textwidth}
  \input{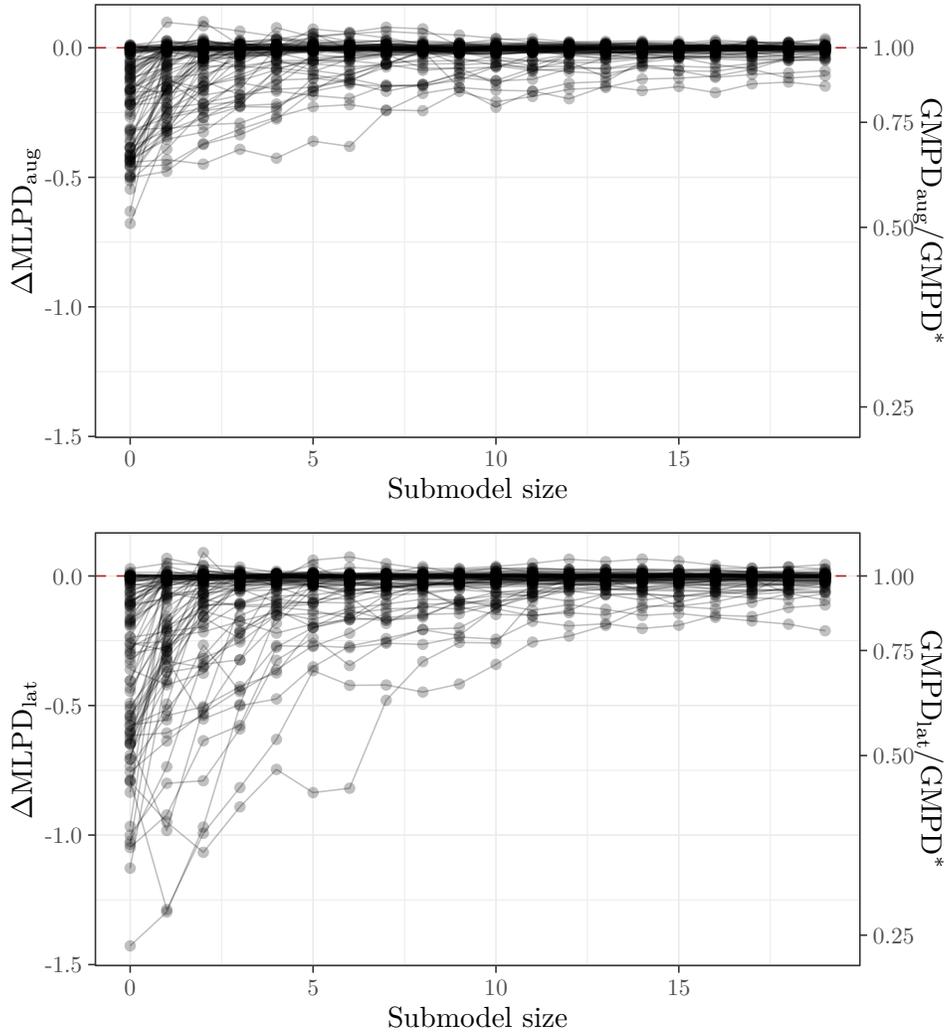}
  \end{adjustbox}
  \caption[Predictive performance relative to the reference model]{
  Relative predictive performance at increasing submodel sizes for the augmented-data projection (top) and the latent projection (bottom) in all $R = 100$ simulation iterations.
  Here, ``relative'' means that the left y-axis shows $\Delta\mathrm{MLPD} = \mathrm{MLPD} - \mathrm{MLPD}^{*}$ (with $\mathrm{MLPD}^{*}$ denoting the reference model MLPD).
  The right y-axis is simply the $\exp(\cdot)$ scale, i.e., it shows $\mathrm{GMPD} / \mathrm{GMPD}^{*}$ (with $\mathrm{GMPD}^{*}$ denoting the reference model GMPD).
  Each line represents one simulation iteration.
  The median $\mathrm{MLPD}^{*}$ across all $R = 100$ simulation iterations is about $-1.4$ (minimum: $-1.7$, first quartile: $-1.6$, third quartile: $-1.1$, maximum: $-0.6$).
  The median $\mathrm{GMPD}^{*}$ across all $R = 100$ simulation iterations is about $0.24$ (minimum: $0.19$, first quartile: $0.20$, third quartile: $0.34$, maximum: $0.54$)
  }
  \label{fig:aug-lat}
\end{figure}

A central part of the \pkg{projpred} workflow is the plot of the chosen performance statistic (relative to the reference model's performance) in dependence of the submodel size.
Basically, this is also what is shown in Figure~\ref{fig:aug-lat}, but slightly adapted to a simulation study:
The lines from all \(R = 100\) simulation iterations are combined into one plot for the augmented-data and the latent projection, respectively.
To avoid an overly crowded plot, the uncertainty bars that are otherwise part of this plot have been omitted.

A reassuring conclusion from Figure~\ref{fig:aug-lat} is that for both projection methods, an increasing submodel size eventually causes the predictive performance of the submodels to approach that of the reference model, although there are simulation iterations where a certain discrepancy to the reference model performance persists even at large submodel sizes.
Nevertheless, we can conclude that both projection methods pass a basic check for being implemented correctly.

\begin{figure}[!tb]
  \centering
  \begin{adjustbox}{max width=\textwidth}
  \input{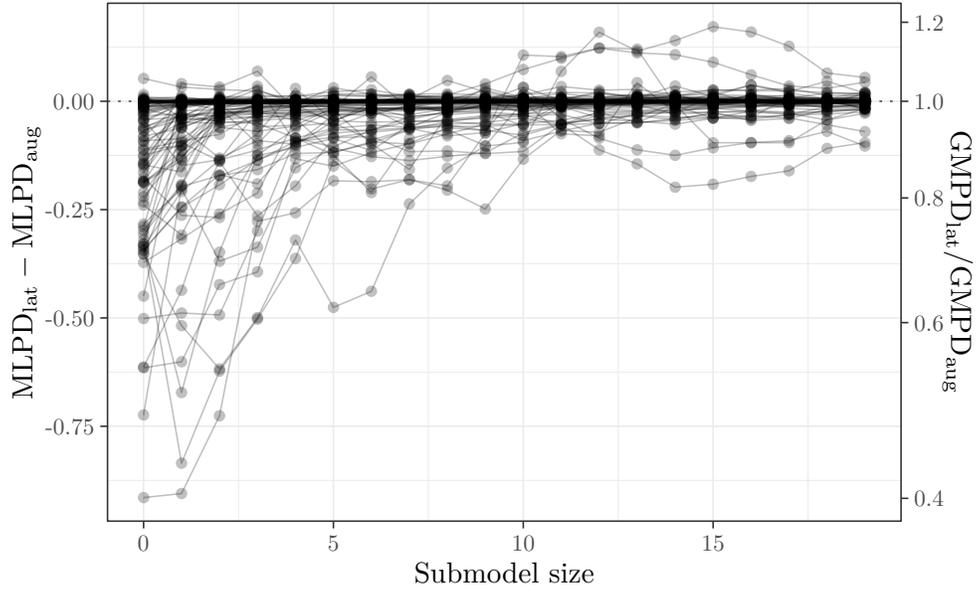}
  \end{adjustbox}
  \caption[Difference of predictive performance between augmented-data and latent projection]{
  Predictive performance based on the latent projection minus predictive performance based on the augmented-data projection, for increasing submodel sizes and all $R = 100$ simulation iterations (represented by lines).
  The right y-axis is simply the $\exp(\cdot)$ scale of the left y-axis.
  Note that for a given simulation iteration, the solution path can differ between the augmented-data and the latent projection
  }
  \label{fig:diff}
\end{figure}

Figure~\ref{fig:aug-lat} also shows that in some simulation iterations, the augmented-data projection's MLPDs at small to moderate submodel sizes are closer to the reference model MLPD than those from the latent projection.
This is even more evident from Figure~\ref{fig:diff} where \(\Delta\mathrm{MLPD}_{\mathrm{lat}} - \Delta\mathrm{MLPD}_{\mathrm{aug}} = \mathrm{MLPD}_{\mathrm{lat}} - \mathrm{MLPD}_{\mathrm{aug}}\) is illustrated.
Figure~\ref{fig:diff} also reveals that there are a few simulation iterations where the latent projection leads to a better predictive performance at large submodel sizes.
These simulation iterations are investigated in more detail in \maxdiff\@.

An inspection of the MLPD (or rather GMPD) values on absolute scale (\resabs) reveals that in extreme cases, the discrepancy in predictive performance between augmented-data and latent projection is indeed non-negligible.

\begin{figure}[!tb]
  \centering
  \begin{adjustbox}{max width=\textwidth}
  \input{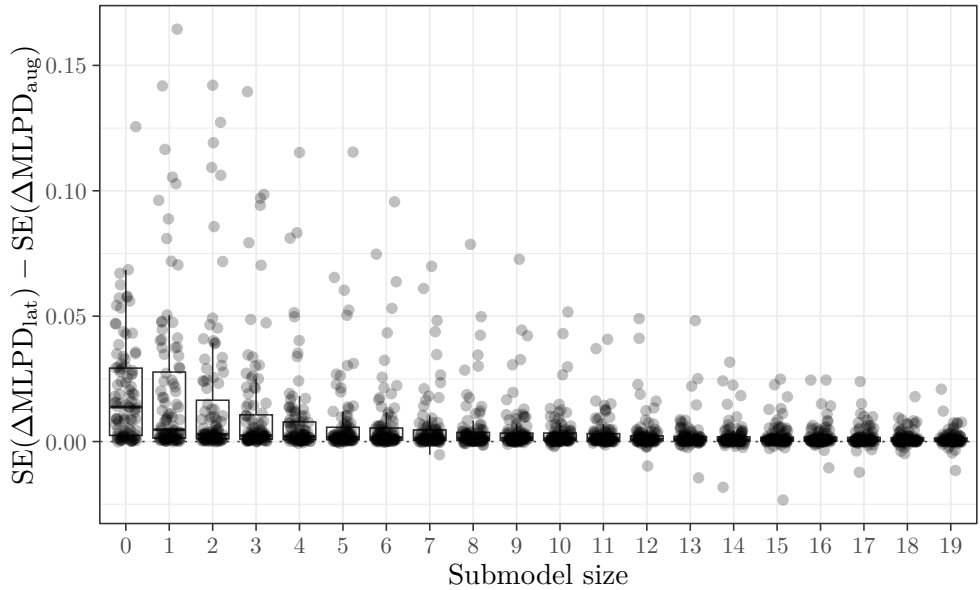}
  \end{adjustbox}
  \caption[Difference of standard error between augmented-data and latent projection]{
  Standard error (SE) in relative predictive performance based on the latent projection minus the same SE based on the augmented-data projection, for increasing submodel sizes and all $R = 100$ simulation iterations (represented by points and summarized by boxplots)
  }
  \label{fig:diff-se}
\end{figure}

The lack of uncertainty bars in Figures~\ref{fig:aug-lat} and~\ref{fig:diff} obscures the fact that all underlying predictive performance values are only \emph{estimates}.
Thus, it is important to inspect, for example, the corresponding standard errors~(SEs).
This is achieved by Figure~\ref{fig:diff-se} which depicts the differences \(\mathrm{SE}(\Delta\mathrm{MLPD}_{\mathrm{lat}}) - \mathrm{SE}(\Delta\mathrm{MLPD}_{\mathrm{aug}})\).
The mostly positive differences in Figure~\ref{fig:diff-se} show that the latent projection is associated with greater uncertainty than the augmented-data projection.
Analogously to the peaks at large submodel sizes from Figure~\ref{fig:diff}, there are latent-projection SEs at large submodel sizes which are noticeably smaller than their counterparts based on the augmented-data projection.
As a side-effect, \maxdiff\ reveals that the SEs from one of the simulation iterations investigated there are part of this rare case.

\begin{figure}[!tb]
  \centering
  \begin{adjustbox}{max width=\textwidth}
\begin{tikzpicture}[x=1pt,y=1pt]
\definecolor{fillColor}{RGB}{255,255,255}
\begin{scope}
\definecolor{drawColor}{RGB}{255,255,255}
\definecolor{fillColor}{RGB}{255,255,255}

\path[draw=drawColor,line width= 0.6pt,line join=round,line cap=round,fill=fillColor] (  0.00,  0.00) rectangle (375.80,232.25);
\end{scope}
\begin{scope}
\definecolor{fillColor}{RGB}{255,255,255}

\path[fill=fillColor] ( 31.71, 30.69) rectangle (370.30,226.75);
\definecolor{drawColor}{gray}{0.92}

\path[draw=drawColor,line width= 0.3pt,line join=round] ( 31.71, 56.10) --
	(370.30, 56.10);

\path[draw=drawColor,line width= 0.3pt,line join=round] ( 31.71, 89.11) --
	(370.30, 89.11);

\path[draw=drawColor,line width= 0.3pt,line join=round] ( 31.71,122.11) --
	(370.30,122.11);

\path[draw=drawColor,line width= 0.3pt,line join=round] ( 31.71,155.12) --
	(370.30,155.12);

\path[draw=drawColor,line width= 0.3pt,line join=round] ( 31.71,188.13) --
	(370.30,188.13);

\path[draw=drawColor,line width= 0.3pt,line join=round] ( 31.71,221.14) --
	(370.30,221.14);

\path[draw=drawColor,line width= 0.6pt,line join=round] ( 31.71, 39.60) --
	(370.30, 39.60);

\path[draw=drawColor,line width= 0.6pt,line join=round] ( 31.71, 72.60) --
	(370.30, 72.60);

\path[draw=drawColor,line width= 0.6pt,line join=round] ( 31.71,105.61) --
	(370.30,105.61);

\path[draw=drawColor,line width= 0.6pt,line join=round] ( 31.71,138.62) --
	(370.30,138.62);

\path[draw=drawColor,line width= 0.6pt,line join=round] ( 31.71,171.63) --
	(370.30,171.63);

\path[draw=drawColor,line width= 0.6pt,line join=round] ( 31.71,204.63) --
	(370.30,204.63);

\path[draw=drawColor,line width= 0.6pt,line join=round] ( 46.02, 30.69) --
	( 46.02,226.75);

\path[draw=drawColor,line width= 0.6pt,line join=round] ( 69.86, 30.69) --
	( 69.86,226.75);

\path[draw=drawColor,line width= 0.6pt,line join=round] ( 93.71, 30.69) --
	( 93.71,226.75);

\path[draw=drawColor,line width= 0.6pt,line join=round] (117.55, 30.69) --
	(117.55,226.75);

\path[draw=drawColor,line width= 0.6pt,line join=round] (141.40, 30.69) --
	(141.40,226.75);

\path[draw=drawColor,line width= 0.6pt,line join=round] (165.24, 30.69) --
	(165.24,226.75);

\path[draw=drawColor,line width= 0.6pt,line join=round] (189.09, 30.69) --
	(189.09,226.75);

\path[draw=drawColor,line width= 0.6pt,line join=round] (212.93, 30.69) --
	(212.93,226.75);

\path[draw=drawColor,line width= 0.6pt,line join=round] (236.77, 30.69) --
	(236.77,226.75);

\path[draw=drawColor,line width= 0.6pt,line join=round] (260.62, 30.69) --
	(260.62,226.75);

\path[draw=drawColor,line width= 0.6pt,line join=round] (284.46, 30.69) --
	(284.46,226.75);

\path[draw=drawColor,line width= 0.6pt,line join=round] (308.31, 30.69) --
	(308.31,226.75);

\path[draw=drawColor,line width= 0.6pt,line join=round] (332.15, 30.69) --
	(332.15,226.75);

\path[draw=drawColor,line width= 0.6pt,line join=round] (356.00, 30.69) --
	(356.00,226.75);
\definecolor{fillColor}{gray}{0.35}

\path[fill=fillColor] ( 35.29, 39.60) rectangle ( 56.75, 42.90);

\path[fill=fillColor] ( 59.13, 39.60) rectangle ( 80.59, 52.80);

\path[fill=fillColor] ( 82.98, 39.60) rectangle (104.44, 62.70);

\path[fill=fillColor] (106.82, 39.60) rectangle (128.28,217.84);

\path[fill=fillColor] (130.67, 39.60) rectangle (152.13, 75.91);

\path[fill=fillColor] (154.51, 39.60) rectangle (175.97, 46.20);

\path[fill=fillColor] (178.36, 39.60) rectangle (199.82, 49.50);

\path[fill=fillColor] (202.20, 39.60) rectangle (223.66, 46.20);

\path[fill=fillColor] (226.04, 39.60) rectangle (247.50, 49.50);

\path[fill=fillColor] (249.89, 39.60) rectangle (271.35, 49.50);

\path[fill=fillColor] (273.73, 39.60) rectangle (295.19, 42.90);

\path[fill=fillColor] (297.58, 39.60) rectangle (319.04, 42.90);

\path[fill=fillColor] (321.42, 39.60) rectangle (342.88, 56.10);

\path[fill=fillColor] (345.27, 39.60) rectangle (366.73, 49.50);
\definecolor{drawColor}{gray}{0.20}

\path[draw=drawColor,line width= 0.6pt,line join=round,line cap=round] ( 31.71, 30.69) rectangle (370.30,226.75);
\end{scope}
\begin{scope}
\definecolor{drawColor}{gray}{0.30}

\node[text=drawColor,anchor=base east,inner sep=0pt, outer sep=0pt, scale=  0.88] at ( 26.76, 36.57) {0};

\node[text=drawColor,anchor=base east,inner sep=0pt, outer sep=0pt, scale=  0.88] at ( 26.76, 69.57) {10};

\node[text=drawColor,anchor=base east,inner sep=0pt, outer sep=0pt, scale=  0.88] at ( 26.76,102.58) {20};

\node[text=drawColor,anchor=base east,inner sep=0pt, outer sep=0pt, scale=  0.88] at ( 26.76,135.59) {30};

\node[text=drawColor,anchor=base east,inner sep=0pt, outer sep=0pt, scale=  0.88] at ( 26.76,168.59) {40};

\node[text=drawColor,anchor=base east,inner sep=0pt, outer sep=0pt, scale=  0.88] at ( 26.76,201.60) {50};
\end{scope}
\begin{scope}
\definecolor{drawColor}{gray}{0.20}

\path[draw=drawColor,line width= 0.6pt,line join=round] ( 28.96, 39.60) --
	( 31.71, 39.60);

\path[draw=drawColor,line width= 0.6pt,line join=round] ( 28.96, 72.60) --
	( 31.71, 72.60);

\path[draw=drawColor,line width= 0.6pt,line join=round] ( 28.96,105.61) --
	( 31.71,105.61);

\path[draw=drawColor,line width= 0.6pt,line join=round] ( 28.96,138.62) --
	( 31.71,138.62);

\path[draw=drawColor,line width= 0.6pt,line join=round] ( 28.96,171.63) --
	( 31.71,171.63);

\path[draw=drawColor,line width= 0.6pt,line join=round] ( 28.96,204.63) --
	( 31.71,204.63);
\end{scope}
\begin{scope}
\definecolor{drawColor}{gray}{0.20}

\path[draw=drawColor,line width= 0.6pt,line join=round] ( 46.02, 27.94) --
	( 46.02, 30.69);

\path[draw=drawColor,line width= 0.6pt,line join=round] ( 69.86, 27.94) --
	( 69.86, 30.69);

\path[draw=drawColor,line width= 0.6pt,line join=round] ( 93.71, 27.94) --
	( 93.71, 30.69);

\path[draw=drawColor,line width= 0.6pt,line join=round] (117.55, 27.94) --
	(117.55, 30.69);

\path[draw=drawColor,line width= 0.6pt,line join=round] (141.40, 27.94) --
	(141.40, 30.69);

\path[draw=drawColor,line width= 0.6pt,line join=round] (165.24, 27.94) --
	(165.24, 30.69);

\path[draw=drawColor,line width= 0.6pt,line join=round] (189.09, 27.94) --
	(189.09, 30.69);

\path[draw=drawColor,line width= 0.6pt,line join=round] (212.93, 27.94) --
	(212.93, 30.69);

\path[draw=drawColor,line width= 0.6pt,line join=round] (236.77, 27.94) --
	(236.77, 30.69);

\path[draw=drawColor,line width= 0.6pt,line join=round] (260.62, 27.94) --
	(260.62, 30.69);

\path[draw=drawColor,line width= 0.6pt,line join=round] (284.46, 27.94) --
	(284.46, 30.69);

\path[draw=drawColor,line width= 0.6pt,line join=round] (308.31, 27.94) --
	(308.31, 30.69);

\path[draw=drawColor,line width= 0.6pt,line join=round] (332.15, 27.94) --
	(332.15, 30.69);

\path[draw=drawColor,line width= 0.6pt,line join=round] (356.00, 27.94) --
	(356.00, 30.69);
\end{scope}
\begin{scope}
\definecolor{drawColor}{gray}{0.30}

\node[text=drawColor,anchor=base,inner sep=0pt, outer sep=0pt, scale=  0.88] at ( 46.02, 19.68) {-3};

\node[text=drawColor,anchor=base,inner sep=0pt, outer sep=0pt, scale=  0.88] at ( 69.86, 19.68) {-2};

\node[text=drawColor,anchor=base,inner sep=0pt, outer sep=0pt, scale=  0.88] at ( 93.71, 19.68) {-1};

\node[text=drawColor,anchor=base,inner sep=0pt, outer sep=0pt, scale=  0.88] at (117.55, 19.68) {0};

\node[text=drawColor,anchor=base,inner sep=0pt, outer sep=0pt, scale=  0.88] at (141.40, 19.68) {1};

\node[text=drawColor,anchor=base,inner sep=0pt, outer sep=0pt, scale=  0.88] at (165.24, 19.68) {2};

\node[text=drawColor,anchor=base,inner sep=0pt, outer sep=0pt, scale=  0.88] at (189.09, 19.68) {3};

\node[text=drawColor,anchor=base,inner sep=0pt, outer sep=0pt, scale=  0.88] at (212.93, 19.68) {4};

\node[text=drawColor,anchor=base,inner sep=0pt, outer sep=0pt, scale=  0.88] at (236.77, 19.68) {5};

\node[text=drawColor,anchor=base,inner sep=0pt, outer sep=0pt, scale=  0.88] at (260.62, 19.68) {7};

\node[text=drawColor,anchor=base,inner sep=0pt, outer sep=0pt, scale=  0.88] at (284.46, 19.68) {13};

\node[text=drawColor,anchor=base,inner sep=0pt, outer sep=0pt, scale=  0.88] at (308.31, 19.68) {$\texttt{NA}_{\mathrm{aug}}$};

\node[text=drawColor,anchor=base,inner sep=0pt, outer sep=0pt, scale=  0.88] at (332.15, 19.68) {$\texttt{NA}_{\mathrm{lat}}$};

\node[text=drawColor,anchor=base,inner sep=0pt, outer sep=0pt, scale=  0.88] at (356.00, 19.68) {$\texttt{NA}_{\mathrm{both}}$};
\end{scope}
\begin{scope}
\definecolor{drawColor}{RGB}{0,0,0}

\node[text=drawColor,anchor=base,inner sep=0pt, outer sep=0pt, scale=  1.10] at (201.01,  7.64) {$G_{\mathrm{lat}} - G_{\mathrm{aug}}$};
\end{scope}
\begin{scope}
\definecolor{drawColor}{RGB}{0,0,0}

\node[text=drawColor,rotate= 90.00,anchor=base,inner sep=0pt, outer sep=0pt, scale=  1.10] at ( 13.08,128.72) {Number of simulation iterations (total: 100)};
\end{scope}
\end{tikzpicture}
  \end{adjustbox}
  \caption[Difference of suggested size between augmented-data and latent projection]{
  Suggested submodel size based on the latent projection ($G_{\mathrm{lat}}$) minus suggested submodel size based on the augmented-data projection ($G_{\mathrm{aug}}$).
  In cases where \texttt{projpred::suggest\_size()} was not able to suggest a size (e.g., because the forward search was terminated before the submodel MLPD could approach the reference model MLPD sufficiently), \texttt{NA} was returned ($\texttt{NA}_{\mathrm{aug}}$: \texttt{NA} for the augmented-data projection only, $\texttt{NA}_{\mathrm{lat}}$: \texttt{NA} for the latent projection only, $\texttt{NA}_{\mathrm{both}}$: \texttt{NA} for the augmented-data projection as well as for the latent projection)
  }
  \label{fig:sgg}
\end{figure}
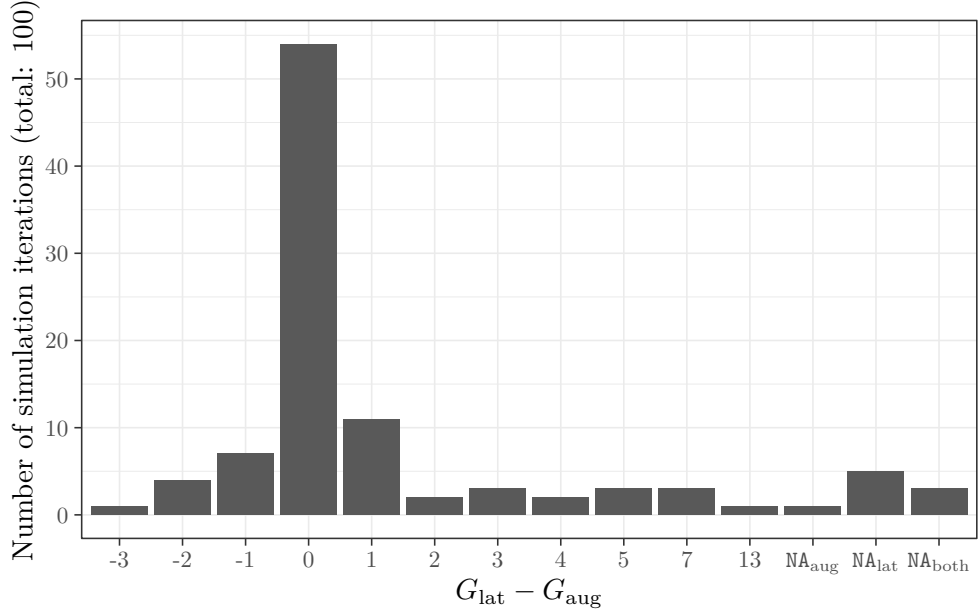

In the typical \pkg{projpred} workflow, the plot of the chosen performance statistic in dependence of the submodel size is mainly used in the decision for a submodel size for the final projection.
Ideally, this plot-based decision is made manually by incorporating subject-matter knowledge, application-specific trade-offs, and the absolute scale of the predictive performance statistic.
In a real-world application, the heuristic offered by \texttt{projpred::suggest\_size()} should only be interpreted as a suggestion, but for the purpose of a simulation study, such a heuristic is helpful.
Figure~\ref{fig:sgg} illustrates the frequency (across the simulation iterations) of all encountered differences \(G_{\mathrm{lat}} - G_{\mathrm{aug}}\) of the sizes \(G_{\mathrm{lat}}\) and \(G_{\mathrm{aug}}\) suggested by this heuristic.
The high peak of the distribution at zero shows that the augmented-data and the latent projection often result in the same suggestion for the submodel size.
Moreover, the slight right-skewness of the distribution (i.e., the presence of a few large positive differences) indicates that there are some simulation iterations where the latent projection leads to a clearly larger suggested size than the augmented-data projection.
This slower convergence of the submodel MLPDs towards the reference model MLPD in case of the latent projection was already visible more directly in Figures~\ref{fig:aug-lat} and~\ref{fig:diff}.
It is also reflected (indirectly) by the larger frequency of \(\texttt{NA}_{\mathrm{lat}}\) compared to \(\texttt{NA}_{\mathrm{aug}}\) in Figure~\ref{fig:sgg}.
A first glance at Figures~\ref{fig:aug-lat} and~\ref{fig:diff} might lead to think that larger suggested sizes in case of the latent projection should be more frequent than they are, but
uncertainty needs to be taken into account, too: The bigger SEs in case of the latent projection (Figure~\ref{fig:diff-se}) may cause the latent projection to arrive at similar suggested sizes as the augmented-data projection, even if the latent-projection submodel MLPDs approach the reference model MLPD more slowly.

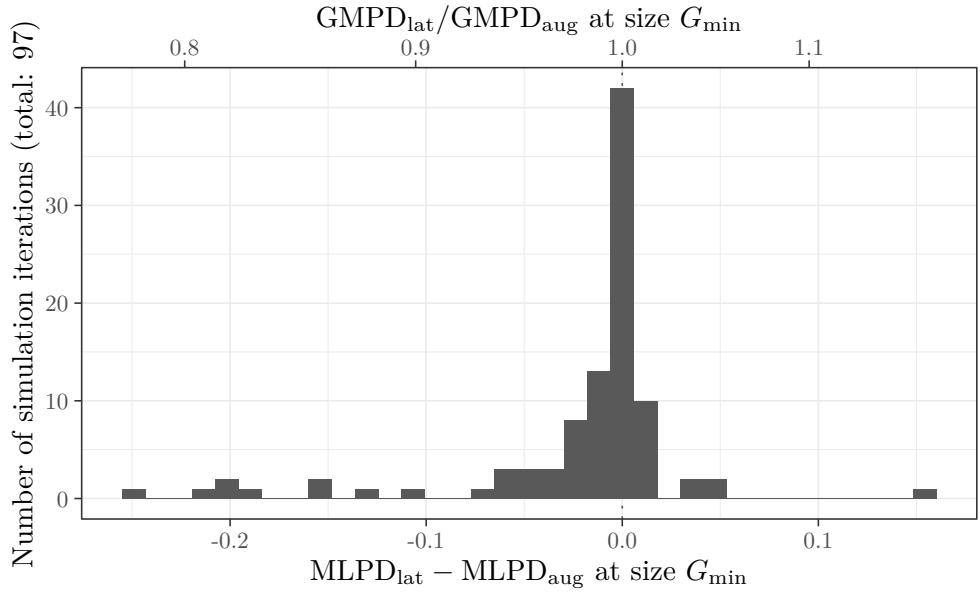
\begin{figure}[!tb]
  \centering
  \begin{adjustbox}{max width=\textwidth}
\begin{tikzpicture}[x=1pt,y=1pt]
\definecolor{fillColor}{RGB}{255,255,255}
\begin{scope}
\definecolor{drawColor}{RGB}{255,255,255}
\definecolor{fillColor}{RGB}{255,255,255}

\path[draw=drawColor,line width= 0.6pt,line join=round,line cap=round,fill=fillColor] (  0.00,  0.00) rectangle (375.80,232.25);
\end{scope}
\begin{scope}
\definecolor{fillColor}{RGB}{255,255,255}

\path[fill=fillColor] ( 31.71, 30.69) rectangle (370.30,201.56);
\definecolor{drawColor}{gray}{0.92}

\path[draw=drawColor,line width= 0.3pt,line join=round] ( 31.71, 56.95) --
	(370.30, 56.95);

\path[draw=drawColor,line width= 0.3pt,line join=round] ( 31.71, 93.93) --
	(370.30, 93.93);

\path[draw=drawColor,line width= 0.3pt,line join=round] ( 31.71,130.92) --
	(370.30,130.92);

\path[draw=drawColor,line width= 0.3pt,line join=round] ( 31.71,167.90) --
	(370.30,167.90);

\path[draw=drawColor,line width= 0.3pt,line join=round] ( 50.76, 30.69) --
	( 50.76,201.56);

\path[draw=drawColor,line width= 0.3pt,line join=round] (124.93, 30.69) --
	(124.93,201.56);

\path[draw=drawColor,line width= 0.3pt,line join=round] (199.10, 30.69) --
	(199.10,201.56);

\path[draw=drawColor,line width= 0.3pt,line join=round] (273.27, 30.69) --
	(273.27,201.56);

\path[draw=drawColor,line width= 0.3pt,line join=round] (347.44, 30.69) --
	(347.44,201.56);

\path[draw=drawColor,line width= 0.6pt,line join=round] ( 31.71, 38.45) --
	(370.30, 38.45);

\path[draw=drawColor,line width= 0.6pt,line join=round] ( 31.71, 75.44) --
	(370.30, 75.44);

\path[draw=drawColor,line width= 0.6pt,line join=round] ( 31.71,112.42) --
	(370.30,112.42);

\path[draw=drawColor,line width= 0.6pt,line join=round] ( 31.71,149.41) --
	(370.30,149.41);

\path[draw=drawColor,line width= 0.6pt,line join=round] ( 31.71,186.40) --
	(370.30,186.40);

\path[draw=drawColor,line width= 0.6pt,line join=round] ( 87.85, 30.69) --
	( 87.85,201.56);

\path[draw=drawColor,line width= 0.6pt,line join=round] (162.02, 30.69) --
	(162.02,201.56);

\path[draw=drawColor,line width= 0.6pt,line join=round] (236.19, 30.69) --
	(236.19,201.56);

\path[draw=drawColor,line width= 0.6pt,line join=round] (310.36, 30.69) --
	(310.36,201.56);
\definecolor{drawColor}{gray}{0.30}

\path[draw=drawColor,line width= 0.6pt,dash pattern=on 1pt off 3pt ,line join=round] (236.19, 30.69) -- (236.19,201.56);
\definecolor{fillColor}{gray}{0.35}

\path[fill=fillColor] ( 47.10, 38.45) rectangle ( 55.90, 42.15);

\path[fill=fillColor] ( 55.90, 38.45) rectangle ( 64.69, 38.45);

\path[fill=fillColor] ( 64.69, 38.45) rectangle ( 73.49, 38.45);

\path[fill=fillColor] ( 73.49, 38.45) rectangle ( 82.28, 42.15);

\path[fill=fillColor] ( 82.28, 38.45) rectangle ( 91.08, 45.85);

\path[fill=fillColor] ( 91.08, 38.45) rectangle ( 99.87, 42.15);

\path[fill=fillColor] ( 99.87, 38.45) rectangle (108.66, 38.45);

\path[fill=fillColor] (108.66, 38.45) rectangle (117.46, 38.45);

\path[fill=fillColor] (117.46, 38.45) rectangle (126.25, 45.85);

\path[fill=fillColor] (126.25, 38.45) rectangle (135.05, 38.45);

\path[fill=fillColor] (135.05, 38.45) rectangle (143.84, 42.15);

\path[fill=fillColor] (143.84, 38.45) rectangle (152.64, 38.45);

\path[fill=fillColor] (152.64, 38.45) rectangle (161.43, 42.15);

\path[fill=fillColor] (161.43, 38.45) rectangle (170.23, 38.45);

\path[fill=fillColor] (170.23, 38.45) rectangle (179.02, 38.45);

\path[fill=fillColor] (179.02, 38.45) rectangle (187.82, 42.15);

\path[fill=fillColor] (187.82, 38.45) rectangle (196.61, 49.55);

\path[fill=fillColor] (196.61, 38.45) rectangle (205.41, 49.55);

\path[fill=fillColor] (205.41, 38.45) rectangle (214.20, 49.55);

\path[fill=fillColor] (214.20, 38.45) rectangle (222.99, 68.04);

\path[fill=fillColor] (222.99, 38.45) rectangle (231.79, 86.53);

\path[fill=fillColor] (231.79, 38.45) rectangle (240.58,193.79);

\path[fill=fillColor] (240.58, 38.45) rectangle (249.38, 75.44);

\path[fill=fillColor] (249.38, 38.45) rectangle (258.17, 38.45);

\path[fill=fillColor] (258.17, 38.45) rectangle (266.97, 45.85);

\path[fill=fillColor] (266.97, 38.45) rectangle (275.76, 45.85);

\path[fill=fillColor] (275.76, 38.45) rectangle (284.56, 38.45);

\path[fill=fillColor] (284.56, 38.45) rectangle (293.35, 38.45);

\path[fill=fillColor] (293.35, 38.45) rectangle (302.15, 38.45);

\path[fill=fillColor] (302.15, 38.45) rectangle (310.94, 38.45);

\path[fill=fillColor] (310.94, 38.45) rectangle (319.74, 38.45);

\path[fill=fillColor] (319.74, 38.45) rectangle (328.53, 38.45);

\path[fill=fillColor] (328.53, 38.45) rectangle (337.32, 38.45);

\path[fill=fillColor] (337.32, 38.45) rectangle (346.12, 38.45);

\path[fill=fillColor] (346.12, 38.45) rectangle (354.91, 42.15);
\definecolor{drawColor}{gray}{0.20}

\path[draw=drawColor,line width= 0.6pt,line join=round,line cap=round] ( 31.71, 30.69) rectangle (370.30,201.56);
\end{scope}
\begin{scope}
\definecolor{drawColor}{gray}{0.30}

\node[text=drawColor,anchor=base,inner sep=0pt, outer sep=0pt, scale=  0.88] at ( 70.68,206.51) {0.8};

\node[text=drawColor,anchor=base,inner sep=0pt, outer sep=0pt, scale=  0.88] at (158.04,206.51) {0.9};

\node[text=drawColor,anchor=base,inner sep=0pt, outer sep=0pt, scale=  0.88] at (236.19,206.51) {1.0};

\node[text=drawColor,anchor=base,inner sep=0pt, outer sep=0pt, scale=  0.88] at (306.88,206.51) {1.1};
\end{scope}
\begin{scope}
\definecolor{drawColor}{gray}{0.20}

\path[draw=drawColor,line width= 0.6pt,line join=round] ( 70.68,201.56) --
	( 70.68,204.31);

\path[draw=drawColor,line width= 0.6pt,line join=round] (158.04,201.56) --
	(158.04,204.31);

\path[draw=drawColor,line width= 0.6pt,line join=round] (236.19,201.56) --
	(236.19,204.31);

\path[draw=drawColor,line width= 0.6pt,line join=round] (306.88,201.56) --
	(306.88,204.31);
\end{scope}
\begin{scope}
\definecolor{drawColor}{gray}{0.30}

\node[text=drawColor,anchor=base east,inner sep=0pt, outer sep=0pt, scale=  0.88] at ( 26.76, 35.42) {0};

\node[text=drawColor,anchor=base east,inner sep=0pt, outer sep=0pt, scale=  0.88] at ( 26.76, 72.41) {10};

\node[text=drawColor,anchor=base east,inner sep=0pt, outer sep=0pt, scale=  0.88] at ( 26.76,109.39) {20};

\node[text=drawColor,anchor=base east,inner sep=0pt, outer sep=0pt, scale=  0.88] at ( 26.76,146.38) {30};

\node[text=drawColor,anchor=base east,inner sep=0pt, outer sep=0pt, scale=  0.88] at ( 26.76,183.37) {40};
\end{scope}
\begin{scope}
\definecolor{drawColor}{gray}{0.20}

\path[draw=drawColor,line width= 0.6pt,line join=round] ( 28.96, 38.45) --
	( 31.71, 38.45);

\path[draw=drawColor,line width= 0.6pt,line join=round] ( 28.96, 75.44) --
	( 31.71, 75.44);

\path[draw=drawColor,line width= 0.6pt,line join=round] ( 28.96,112.42) --
	( 31.71,112.42);

\path[draw=drawColor,line width= 0.6pt,line join=round] ( 28.96,149.41) --
	( 31.71,149.41);

\path[draw=drawColor,line width= 0.6pt,line join=round] ( 28.96,186.40) --
	( 31.71,186.40);
\end{scope}
\begin{scope}
\definecolor{drawColor}{gray}{0.20}

\path[draw=drawColor,line width= 0.6pt,line join=round] ( 87.85, 27.94) --
	( 87.85, 30.69);

\path[draw=drawColor,line width= 0.6pt,line join=round] (162.02, 27.94) --
	(162.02, 30.69);

\path[draw=drawColor,line width= 0.6pt,line join=round] (236.19, 27.94) --
	(236.19, 30.69);

\path[draw=drawColor,line width= 0.6pt,line join=round] (310.36, 27.94) --
	(310.36, 30.69);
\end{scope}
\begin{scope}
\definecolor{drawColor}{gray}{0.30}

\node[text=drawColor,anchor=base,inner sep=0pt, outer sep=0pt, scale=  0.88] at ( 87.85, 19.68) {-0.2};

\node[text=drawColor,anchor=base,inner sep=0pt, outer sep=0pt, scale=  0.88] at (162.02, 19.68) {-0.1};

\node[text=drawColor,anchor=base,inner sep=0pt, outer sep=0pt, scale=  0.88] at (236.19, 19.68) {0.0};

\node[text=drawColor,anchor=base,inner sep=0pt, outer sep=0pt, scale=  0.88] at (310.36, 19.68) {0.1};
\end{scope}
\begin{scope}
\definecolor{drawColor}{RGB}{0,0,0}

\node[text=drawColor,anchor=base,inner sep=0pt, outer sep=0pt, scale=  1.10] at (201.01,217.03) {$\mathrm{GMPD}_{\mathrm{lat}} / \mathrm{GMPD}_{\mathrm{aug}}$ at size $G_{\mathrm{min}}$};
\end{scope}
\begin{scope}
\definecolor{drawColor}{RGB}{0,0,0}

\node[text=drawColor,anchor=base,inner sep=0pt, outer sep=0pt, scale=  1.10] at (201.01,  7.64) {$\mathrm{MLPD}_{\mathrm{lat}} - \mathrm{MLPD}_{\mathrm{aug}}$ at size $G_{\mathrm{min}}$};
\end{scope}
\begin{scope}
\definecolor{drawColor}{RGB}{0,0,0}

\node[text=drawColor,rotate= 90.00,anchor=base,inner sep=0pt, outer sep=0pt, scale=  1.10] at ( 13.08,116.12) {Number of simulation iterations (total: 97)};
\end{scope}
\end{tikzpicture}
  \end{adjustbox}
  \caption[Difference of predictive performance between augmented-data and latent projection at the minimum of the two suggested sizes]{
  Predictive performance based on the latent projection minus predictive performance based on the augmented-data projection, at size $G_{\mathrm{min}} = \min(G_{\mathrm{aug}}, G_{\mathrm{lat}})$.
  The total of $97$ instead of $R = 100$ simulation iterations is caused by $3$ iterations where both projection methods resulted in a suggested size of \texttt{NA}.
  The top x-axis is simply the $\exp(\cdot)$ scale of the bottom x-axis
  }
  \label{fig:diff-sgg}
\end{figure}

The slower convergence towards the reference model MLPD in case of the latent projection is also visible in a slight left-skewness (with peak around zero) of the distribution of \(\mathrm{MLPD}_{\mathrm{lat}} - \mathrm{MLPD}_{\mathrm{aug}}\) at submodel size \(G_{\mathrm{min}} = \min(G_{\mathrm{aug}}, G_{\mathrm{lat}})\) (provided at least one of \(G_{\mathrm{aug}}\) and \(G_{\mathrm{lat}}\) is non-\texttt{NA}) across all simulation iterations (Figure~\ref{fig:diff-sgg}).

\begin{figure}[!tb]
  \centering
  \input{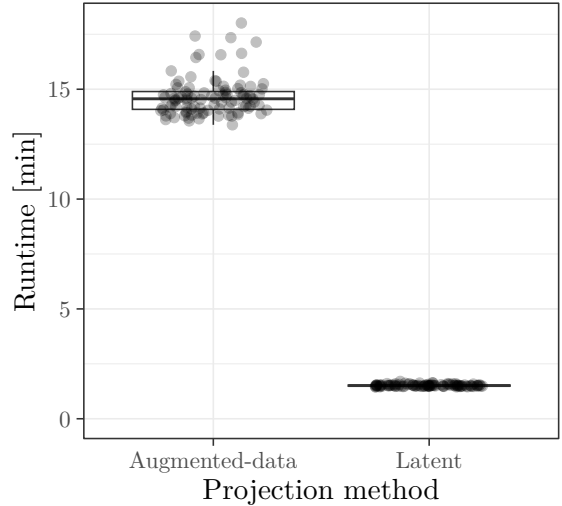}
  \caption[Runtime]{
  Runtime (in minutes) of \texttt{projpred::varsel()} based on the augmented-data and the latent projection, for all $R = 100$ simulation iterations (represented by points and summarized by boxplots)
  }
  \label{fig:time}
\end{figure}

Finally, Figure~\ref{fig:time} shows the runtime of the \texttt{projpred::varsel()} call for both projection methods.
Clearly, the augmented-data projection takes much longer (median runtime across all simulation iterations: ca.\ 14.6 minutes) than the latent projection (median runtime across all simulation iterations: ca.\ 1.5 minutes).
This is the price to pay for the exact projection instead of the approximate latent projection.

\section{Example: Renal cell carcinoma subtyping}\label{ex}

We illustrate the application of the augmented-data projection embedded in a projection predictive variable selection for a nominal response variable using a cancer dataset from the Institute of Pathology of the Rostock University Medical Center (Germany).
This dataset
consists of those 285 observations (patients) with complete records from the larger dataset used by \citet{zimpfer_histopathologische_2019}.

\citet{zimpfer_histopathologische_2019} conducted a retrospective study for renal cell carcinoma~(RCC) subtyping in accordance with the 2016 WHO classification.
RCC subtyping is of prognostic relevance for patients and thus crucial to be determined accurately.
In \citet{zimpfer_histopathologische_2019}, RCC subtyping was performed histologically by trained pathologists.
Our data contains three RCC subtypes: clear-cell RCC (relative frequency: ca.\ \(86.0\,\%\)), papillary RCC (ca.\ \(9.5\,\%\)), and a set of rare (WHO-unclassified) subtypes (ca.\ \(4.6\,\%\)).

Despite the focus on determining the RCC subtype accurately, it is also helpful to predict the RCC subtype as early as possible during the process of patient care.
Thus, we apply a projection predictive variable selection with the three-level RCC subtype as response.
On the side of the predictors, our reference model consists of the main effects and all possible two-way interactions of the following seven predictor variables which were chosen based on Table 2 of \citet{zimpfer_histopathologische_2019}:

\begin{itemize}
\item
  \texttt{age}: age at diagnosis (in years),
\item
  \texttt{sex}: sex (\texttt{"female"} or \texttt{"male"}),
\item
  \texttt{grade}: histologic tumor grade (coded as \texttt{"G1G2"} for grades G1--G2 and \texttt{"G3G4"} for G3--G4),
\item
  \texttt{stage}: histologic tumor stage (coded as \texttt{"T1T2"} for stages T1--T2 and \texttt{"T3T4"} for T3--T4),
\item
  \texttt{nodes}: nodal metastases spread nearby (coded as \texttt{"no"} for N0 and \texttt{"yes"} for N1),
\item
  \texttt{metastases}: metastases 0-6 months post-diagnosis (coded as \texttt{"no"} for M0 and \texttt{"yes"} for M1),
\item
  \texttt{resection}: classification of the resection margin (coded as \texttt{"R0"} for R0 and \texttt{"R1R2"} for R1--R2).
\end{itemize}

In the following, we only describe modeling choices deviating from the defaults of the respective R function arguments.

For fitting the reference model, we use the \texttt{brms::categorical()} response family from the R package \pkg{brms}.
For the regression coefficients, we choose the R2-D2 prior \citep{zhang_bayesian_2022} as implemented in \pkg{brms}.
In case of the \texttt{brms::categorical()} family, the R2-D2 prior's \(R^2\) parameter does not have an intuitive interpretation (in contrast to normal linear models), but
smaller \(R^2\) values still imply a stronger penalization.
Here, we choose a mean of \(0.4\) and a pseudo-precision parameter of \(2.5\) for the Beta prior on \(R^2\), so slightly more penalization than implied by the default uniform Beta prior.
The current implementation of the R2-D2 prior in \pkg{brms} requires a comparable scale of the predictors (except if differing scales have a meaning with respect to the relevance of predictors, in the sense that predictors with a larger scale should be more relevant, which we don't assume here). Thus, as suggested by \citet{gelman_weakly_2008}, we scale the only continuous predictor variable \texttt{age} to a standard deviation of \(0.5\) (which corresponds to the standard deviation of a binary predictor with a relative frequency of \(50\,\%\) for both categories).
Prior to scaling \texttt{age}, we center it to a mean of \(0\).

The convergence of the Markov chains in the \pkg{brms} reference model fit
seems to be given: All checks that we already performed in the simulation study (section~\ref{sim-setup}) are passed.
Furthermore, we conduct some basic checks for the reference model to be appropriate from a predictive point of view.
These checks
(not shown here)
reveal that the reference model's predictions are largely driven by the intercepts.
(In a \texttt{brms::categorical()} model, the intercepts transformed to response scale---i.e., to probabilities---reflect the hypothetical frequencies of the response categories at predictor values of zero.)
In this sense, the reference model (or rather the data it is based upon) is suboptimal, but still sufficient for illustrative purposes.

Within \pkg{projpred}, we perform the projection predictive variable selection using a \(K\)-fold cross-validation (\(K\)-fold CV), here with \(K = 30\).
Based on a preliminary \texttt{projpred::cv\_varsel()} run with Pareto-smoothed importance sampling leave-one-out CV \citep[PSIS-LOO CV,][]{vehtari_practical_2017, vehtari_pareto_2022} and a full-data search (i.e., a search that was not run separately for each CV fold), we restrict the maximum submodel size for the fold-wise searches in the final \texttt{projpred::cv\_varsel()} run (the \(K\)-fold one) to \(3\), thereby saving computational resources.

\begin{figure}[!tb]
  \centering
  \begin{adjustbox}{max width=\textwidth}
\begin{tikzpicture}[x=1pt,y=1pt]
\definecolor{fillColor}{RGB}{255,255,255}
\begin{scope}
\definecolor{drawColor}{RGB}{255,255,255}
\definecolor{fillColor}{RGB}{255,255,255}

\path[draw=drawColor,line width= 0.6pt,line join=round,line cap=round,fill=fillColor] (  0.00,  0.00) rectangle (375.80,232.25);
\end{scope}
\begin{scope}
\definecolor{fillColor}{RGB}{255,255,255}

\path[fill=fillColor] ( 41.49, 30.69) rectangle (337.25,226.75);
\definecolor{drawColor}{gray}{0.92}

\path[draw=drawColor,line width= 0.3pt,line join=round] ( 41.49, 58.26) --
	(337.25, 58.26);

\path[draw=drawColor,line width= 0.3pt,line join=round] ( 41.49,113.76) --
	(337.25,113.76);

\path[draw=drawColor,line width= 0.3pt,line join=round] ( 41.49,169.27) --
	(337.25,169.27);

\path[draw=drawColor,line width= 0.3pt,line join=round] ( 41.49,224.77) --
	(337.25,224.77);

\path[draw=drawColor,line width= 0.6pt,line join=round] ( 41.49, 86.01) --
	(337.25, 86.01);

\path[draw=drawColor,line width= 0.6pt,line join=round] ( 41.49,141.52) --
	(337.25,141.52);

\path[draw=drawColor,line width= 0.6pt,line join=round] ( 41.49,197.02) --
	(337.25,197.02);

\path[draw=drawColor,line width= 0.6pt,line join=round] ( 54.93, 30.69) --
	( 54.93,226.75);

\path[draw=drawColor,line width= 0.6pt,line join=round] (144.56, 30.69) --
	(144.56,226.75);

\path[draw=drawColor,line width= 0.6pt,line join=round] (234.18, 30.69) --
	(234.18,226.75);

\path[draw=drawColor,line width= 0.6pt,line join=round] (323.81, 30.69) --
	(323.81,226.75);
\definecolor{drawColor}{RGB}{139,0,0}

\path[draw=drawColor,line width= 0.6pt,dash pattern=on 4pt off 4pt ,line join=round] ( 41.49,197.02) -- (337.25,197.02);
\definecolor{drawColor}{RGB}{0,0,0}

\path[draw=drawColor,draw opacity=0.55,line width= 0.6pt,line join=round] ( 54.93, 39.60) -- ( 54.93,139.60);

\path[draw=drawColor,draw opacity=0.55,line width= 0.6pt,line join=round] (144.56,168.06) -- (144.56,206.17);

\path[draw=drawColor,draw opacity=0.55,line width= 0.6pt,line join=round] (234.18,184.93) -- (234.18,217.84);

\path[draw=drawColor,draw opacity=0.55,line width= 0.6pt,line join=round] (323.81,172.34) -- (323.81,194.90);
\definecolor{drawColor}{RGB}{0,0,0}

\path[draw=drawColor,line width= 0.6pt,line join=round] ( 54.93, 89.60) --
	(144.56,187.11) --
	(234.18,201.38) --
	(323.81,183.62);
\definecolor{fillColor}{RGB}{0,0,0}

\path[draw=drawColor,line width= 0.4pt,line join=round,line cap=round,fill=fillColor] ( 54.93, 89.60) circle (  1.96);

\path[draw=drawColor,line width= 0.4pt,line join=round,line cap=round,fill=fillColor] (144.56,187.11) circle (  1.96);

\path[draw=drawColor,line width= 0.4pt,line join=round,line cap=round,fill=fillColor] (234.18,201.38) circle (  1.96);

\path[draw=drawColor,line width= 0.4pt,line join=round,line cap=round,fill=fillColor] (323.81,183.62) circle (  1.96);
\definecolor{drawColor}{gray}{0.20}

\path[draw=drawColor,line width= 0.6pt,line join=round,line cap=round] ( 41.49, 30.69) rectangle (337.25,226.75);
\end{scope}
\begin{scope}
\definecolor{drawColor}{gray}{0.30}

\node[text=drawColor,anchor=base east,inner sep=0pt, outer sep=0pt, scale=  0.88] at ( 36.54, 82.98) {-0.04};

\node[text=drawColor,anchor=base east,inner sep=0pt, outer sep=0pt, scale=  0.88] at ( 36.54,138.49) {-0.02};

\node[text=drawColor,anchor=base east,inner sep=0pt, outer sep=0pt, scale=  0.88] at ( 36.54,193.99) {0.00};
\end{scope}
\begin{scope}
\definecolor{drawColor}{gray}{0.20}

\path[draw=drawColor,line width= 0.6pt,line join=round] ( 38.74, 86.01) --
	( 41.49, 86.01);

\path[draw=drawColor,line width= 0.6pt,line join=round] ( 38.74,141.52) --
	( 41.49,141.52);

\path[draw=drawColor,line width= 0.6pt,line join=round] ( 38.74,197.02) --
	( 41.49,197.02);
\end{scope}
\begin{scope}
\definecolor{drawColor}{gray}{0.20}

\path[draw=drawColor,line width= 0.6pt,line join=round] (337.25, 83.73) --
	(340.00, 83.73);

\path[draw=drawColor,line width= 0.6pt,line join=round] (337.25,140.95) --
	(340.00,140.95);

\path[draw=drawColor,line width= 0.6pt,line join=round] (337.25,197.02) --
	(340.00,197.02);
\end{scope}
\begin{scope}
\definecolor{drawColor}{gray}{0.30}

\node[text=drawColor,anchor=base west,inner sep=0pt, outer sep=0pt, scale=  0.88] at (342.20, 80.70) {0.96};

\node[text=drawColor,anchor=base west,inner sep=0pt, outer sep=0pt, scale=  0.88] at (342.20,137.92) {0.98};

\node[text=drawColor,anchor=base west,inner sep=0pt, outer sep=0pt, scale=  0.88] at (342.20,193.99) {1.00};
\end{scope}
\begin{scope}
\definecolor{drawColor}{gray}{0.20}

\path[draw=drawColor,line width= 0.6pt,line join=round] ( 54.93, 27.94) --
	( 54.93, 30.69);

\path[draw=drawColor,line width= 0.6pt,line join=round] (144.56, 27.94) --
	(144.56, 30.69);

\path[draw=drawColor,line width= 0.6pt,line join=round] (234.18, 27.94) --
	(234.18, 30.69);

\path[draw=drawColor,line width= 0.6pt,line join=round] (323.81, 27.94) --
	(323.81, 30.69);
\end{scope}
\begin{scope}
\definecolor{drawColor}{gray}{0.30}

\node[text=drawColor,anchor=base,inner sep=0pt, outer sep=0pt, scale=  0.88] at ( 54.93, 19.68) {0};

\node[text=drawColor,anchor=base,inner sep=0pt, outer sep=0pt, scale=  0.88] at (144.56, 19.68) {1};

\node[text=drawColor,anchor=base,inner sep=0pt, outer sep=0pt, scale=  0.88] at (234.18, 19.68) {2};

\node[text=drawColor,anchor=base,inner sep=0pt, outer sep=0pt, scale=  0.88] at (323.81, 19.68) {3};
\end{scope}
\begin{scope}
\definecolor{drawColor}{RGB}{0,0,0}

\node[text=drawColor,anchor=base,inner sep=0pt, outer sep=0pt, scale=  1.10] at (189.37,  7.64) {Submodel size (number of predictor terms)};
\end{scope}
\begin{scope}
\definecolor{drawColor}{RGB}{0,0,0}

\node[text=drawColor,rotate= 90.00,anchor=base,inner sep=0pt, outer sep=0pt, scale=  1.10] at ( 13.08,128.72) {$\Delta\mathrm{MLPD}$};
\end{scope}
\begin{scope}
\definecolor{drawColor}{RGB}{0,0,0}

\node[text=drawColor,rotate=-90.00,anchor=base,inner sep=0pt, outer sep=0pt, scale=  1.10] at (360.59,128.72) {$\mathrm{GMPD} / \mathrm{GMPD}^{*}$};
\end{scope}
\end{tikzpicture}
  \end{adjustbox}
  \caption[Predictive performance relative to the reference model]{
  Relative predictive performance for increasing submodel sizes in the RCC example from section~\ref{ex}.
  Here, ``relative'' means that the left y-axis shows $\Delta\mathrm{MLPD} = \mathrm{MLPD} - \mathrm{MLPD}^{*}$ (with $\mathrm{MLPD}^{*}$ denoting the reference model MLPD).
  The right y-axis is simply the $\exp(\cdot)$ scale, i.e., it shows $\mathrm{GMPD} / \mathrm{GMPD}^{*}$ (with $\mathrm{GMPD}^{*}$ denoting the reference model GMPD).
  The vertical uncertainty bars indicate $\mathrm{SE}(\Delta\mathrm{MLPD})$ (one such SE to either side of the point estimate)
  }
  \label{fig:ex-perf-plot}
\end{figure}
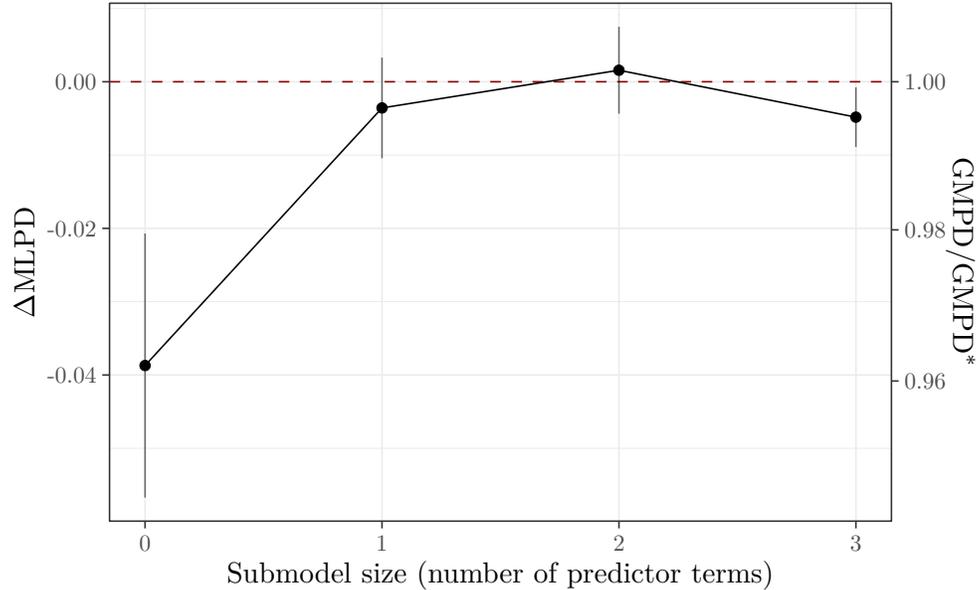

The whole \pkg{projpred} part of our code takes approximately 15 minutes on a standard desktop machine. The final \texttt{projpred::cv\_varsel()} run yields the predictive performance plot depicted in Figure~\ref{fig:ex-perf-plot}.

Based on Figure~\ref{fig:ex-perf-plot}, we choose a submodel size of \(2\).
The heuristic implemented in \texttt{projpred::suggest\_size()} would have given a size of \(1\) (because size \(1\) is the smallest size where the submodel MLPD point estimate is less than one standard error smaller than the reference model MLPD point estimate).
Here, we choose the slightly bigger size of \(2\) due to the special medical context where
the primary goal is predictive accuracy, and sparsity being a secondary goal.

The summary of the fold-wise solution paths presented in Table~\ref{tab:ex-solpaths-cv} shows that all \(K = 30\) CV folds agree on the first two predictors: \texttt{metastases} and \texttt{nodes} (in this order).
Thus, our selected submodel consists of these two predictors.
After a final projection of the reference model onto this submodel (this time using the draw-by-draw method, i.e., projecting each posterior draw from the reference model onto the submodel parameter space without any clustering), we can make predictions with this submodel.
These predictions are presented in Table~\ref{tab:ex-preds} (this compact form is possible here because there are only two binary predictors).
Although the absolute changes in the predictive probabilities might at first seem quite large (up to about \(22\,\%\) when changing only one predictor at a time, and up to about \(27\,\%\) when changing both predictors simultaneously), the predictive probabilities are still dominated by the empirical frequencies of the response categories in the data and thus by the intercepts.
As mentioned above, this was already observable in the reference model.
Therefore, it is clear that this pattern is also visible here:
Model selection cannot be expected to yield a model with better predictions than the reference model \citep{vehtari_survey_2012, piironen_comparison_2017}, especially in the context of projections which are essentially fitting to the fit of the reference model.

\begin{table}[!tb]
\centering
\caption[Summary of the fold-wise solution paths]{Summary of the fold-wise solution paths. For a given submodel size and a given predictor term from the full-data solution path, the table shows the proportion of CV folds that have this predictor term at this size in their solution path. Only submodel sizes $1$ to $3$ are shown because the search was terminated intentionally at size $3$ (see text)}
\label{tab:ex-solpaths-cv}
\begin{tabular}{rrrr}
  \toprule
Submodel size & \texttt{metastases} & \texttt{nodes} & \texttt{grade} \\
  \midrule
1 & $100\,\%$ & $0\,\%$ & $0\,\%$ \\
  2 & $0\,\%$ & $100\,\%$ & $0\,\%$ \\
  3 & $0\,\%$ & $0\,\%$ & $63\,\%$ \\
   \bottomrule
\end{tabular}
\end{table}

\begin{table}[!tb]
\centering
\caption[Predictions from the final submodel]{All possible predictions from the final submodel. Here, the predictions are probabilities for the categories of the response, the RCC subtype. The ``rare'' RCCs are the WHO-unclassified ones}
\label{tab:ex-preds}
\begin{tabular}{lrrr}
  \toprule
 & \multicolumn{3}{c}{RCC subtype} \\
Predictor combination & Clear-cell & Papillary & Rare \\
  \midrule
\texttt{metastases} = \texttt{"no"}, \texttt{nodes} = \texttt{"no"} & $88\,\%$ & $10\,\%$ & $2\,\%$ \\
  \texttt{metastases} = \texttt{"yes"}, \texttt{nodes} = \texttt{"no"} & $79\,\%$ & $9\,\%$ & $12\,\%$ \\
  \texttt{metastases} = \texttt{"no"}, \texttt{nodes} = \texttt{"yes"} & $83\,\%$ & $11\,\%$ & $7\,\%$ \\
  \texttt{metastases} = \texttt{"yes"}, \texttt{nodes} = \texttt{"yes"} & $63\,\%$ & $8\,\%$ & $29\,\%$ \\
   \bottomrule
\end{tabular}
\end{table}

\section{Discussion}\label{disc}

We have presented how the projective part of the projection predictive variable selection can be performed in case of a discrete response family with finite support.
This augmented-data projection has been implemented as an extension of the \pkg{projpred} R package.

Apart from the presentation of the methodology, the purpose of this paper was to compare the augmented-data projection to the latent projection, an alternative projection method that is far more general than the augmented-data projection and covers many discrete finite-support response families as well.
The simulation study we have conducted to this end demonstrated that most of the time, the two projection methods behave quite similarly in terms of predictive performance and the submodel size found by the \texttt{projpred::suggest\_size()} heuristic.
In some cases, the augmented-data projection yields a better predictive performance and (although not necessarily in the same cases) a smaller suggested size than the latent projection.
In even less frequent cases, it is the latent projection which yields a better predictive performance and a smaller suggested size.

Overall (i.e., across all simulation iterations), the predictive performance of the submodels and the variable selection based upon it seem to be more stable in case of the augmented-data projection.
This is probably due to the exact nature of the augmented-data projection, as opposed to the approximate nature of the latent projection.
For example, in case of the ordinal family used here, one reason for the worse stability of the latent projection could be that it uses the reference model's draws of the threshold parameters to compute response-scale output (such as the response-scale MLPD) for a submodel:
In general, the smaller the submodel size, the larger the lack of fit between the latent predictor of a submodel and the latent predictor of the reference model will be.
When using an ad-hoc solution for computing (response-scale) predictive probabilities by relying on the reference model's thresholds, a lack of fit in the latent predictor causes the predictive probabilities of a submodel to become suboptimal without the projection noticing this (and thus without the possibility for the projection to adjust the regression coefficients).
In contrast, the augmented-data projection aims at reproducing directly the predictive probabilities of the reference model, adjusting both, the regression coefficients and the thresholds of a submodel.
In principle, the latent projection also allows to calculate the predictive performance statistic(s) and other post-projection quantities on latent scale.
By converting the results from the augmented-data projection to latent scale as well, we could have tried to compare the augmented-data and the latent projection on latent scale.
However, in settings like ours where there is an independent test dataset (and the same applies to \(K\)-fold CV), it is not straightforward to define how the latent-scale predictions for the test dataset should be calculated (using the reference model fit based on the training data would induce a dependency between training and test data).
Furthermore, latent-scale performance statistics like the latent-scale MLPD are not easily interpretable.
Hence, we did not perform latent-scale analyses in our simulation study.

MLPD was the only predictive performance statistic in our simulation study.
In principle, the classification accuracy could be used as an alternative performance statistic in discrete finite-support observation models.
However, especially in case of a moderate to large number of response categories (like the \(J = 5\) categories in our simulation study), this comes with a loss of information that MLPD does not exhibit:
For example, if the true response category of an observation is category \(3\) (out of \(5\)) and a model gives a predictive probability of \(23\,\%\) for category \(3\), a predictive probability of \(24\,\%\) for category \(4\), and predictive probabilities smaller than \(23\,\%\) for all other categories, then the prediction of the highest-probability category would lead to a misclassification in the zero-one utility spirit of the classification accuracy.
MLPD is smoother in the sense that the log predictive probability of that observation is \(\log(0.23)\), which would not differ much from the log predictive probability of \(\log(0.24)\) in a situation where the predictive probabilities for categories \(3\) and \(4\) were reversed.
In any case, even if the accuracy may be considered appropriate in some use cases (after all, the choice of performance statistic is an application-specific one), we do not expect our main conclusions to change significantly in case of alternative performance statistics.

The cost of the augmented-data projection's higher stability is a considerable increase in runtime.
Because of this, it might be helpful to use the latent projection for preliminary results in the model-building workflow and to use the augmented-data projection afterwards for final results.
One particular purpose of a preliminary latent-projection run could be to find a reasonable value for argument \texttt{nterms\_max} of \texttt{projpred::varsel()} or \texttt{projpred::cv\_varsel()} (this argument determines up to which submodel size the search should be conducted) because often, \texttt{nterms\_max} can be chosen smaller than the value implied by the default heuristic, which reduces the runtime for the final augmented-data projection significantly.

An advantage of the augmented-data projection that was shortly mentioned in section~\ref{math-finite} and later illustrated in the example from section~\ref{ex} is the support for nominal families like \texttt{brms::categorical()}.
So far, such families are not supported by the latent projection.

In the future (and if requested by users), the implementation of the augmented-data projection in \pkg{projpred} can be extended to more exotic discrete finite-support response families in a straightforward manner (see section~\ref{math-finite}).

Furthermore, the augmented-data projection might also be applicable to continuous response families and discrete families with infinite support, using either a Monte Carlo or a discretization approach for achieving an artificial support that is discrete and finite. The Monte Carlo approach might require a clustering or some other kind of grouping of the response draws to arrive at a practicable number of response categories. For the discretization approach, it might be possible to borrow ideas from \citet{rover_discrete_2017}.

Finally, we note that the augmented-data projection in \pkg{projpred} also supports multilevel models.
Since the projection predictive variable selection for multilevel models (in general) is currently subject to more detailed investigations, we leave the comparison of augmented-data and latent projection for multilevel models for future research.

%

\section{Acknowledgments}

We thank the Academy of Finland (grant 340721) for partial funding of this research.
We also acknowledge the computational resources provided by the University of Rostock.

\begin{appendices}
\counterwithin{table}{section}
\counterwithin{figure}{section}
\counterwithin{equation}{section}

\section{Simulation iterations with better predictive performance under the latent projection}\label{maxdiff}

\begin{figure}[!tb]
  \centering
  \begin{adjustbox}{max width=\textwidth}
  \input{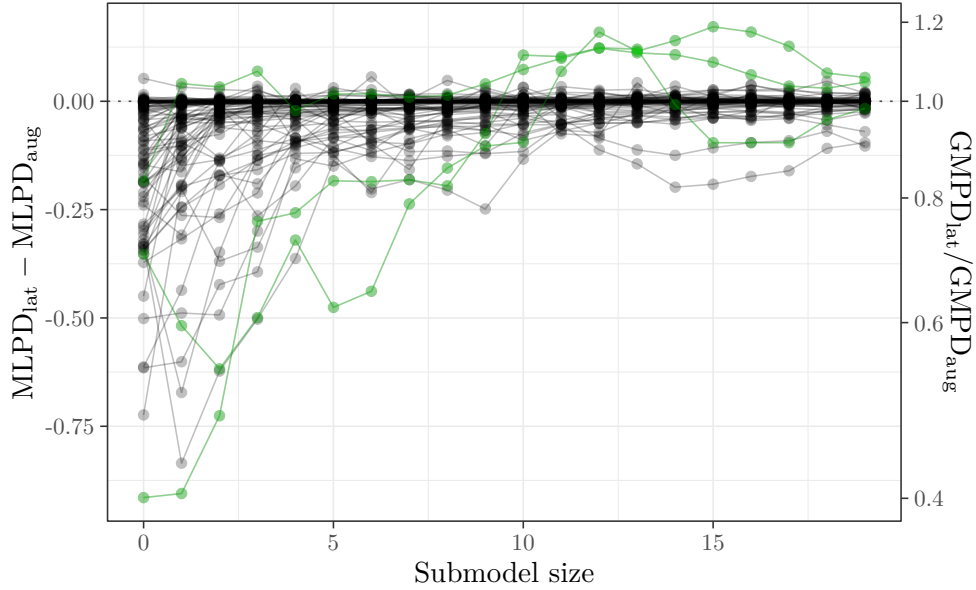}
  \end{adjustbox}
  \caption[Difference of predictive performance between augmented-data and latent projection in simulation iterations with better predictive performance under the latent projection]{
  Predictive performance difference between augmented-data and latent projection in the three simulation iterations ($31$, $75$, and $18$; highlighted in green) with the largest values for $\mathrm{MLPD}_{\mathrm{lat}} - \mathrm{MLPD}_{\mathrm{aug}}$.
  Apart from the highlighting, this is the same as \figdiff
  }
  \label{fig:diff-maxdiff}
\end{figure}

Figure~\ref{fig:diff-maxdiff} is the same as \figdiff, but with three simulation iterations highlighted, namely those with the largest values for \(\mathrm{MLPD}_{\mathrm{lat}} - \mathrm{MLPD}_{\mathrm{aug}}\) across all submodel sizes that the forward search runs through (i.e., the three iterations where the MLPD advantage of the latent projection compared to the augmented-data projection is the largest, no matter at which submodel size).
These three simulation iterations are the \(31^{\mathrm{st}}\), the \(75^{\mathrm{th}}\), and the \(18^{\mathrm{th}}\) (sorted from largest \(\mathrm{MLPD}_{\mathrm{lat}} - \mathrm{MLPD}_{\mathrm{aug}}\) to smallest).

\begin{figure}[!tb]
  \centering
  \begin{adjustbox}{max width=\textwidth}
  \input{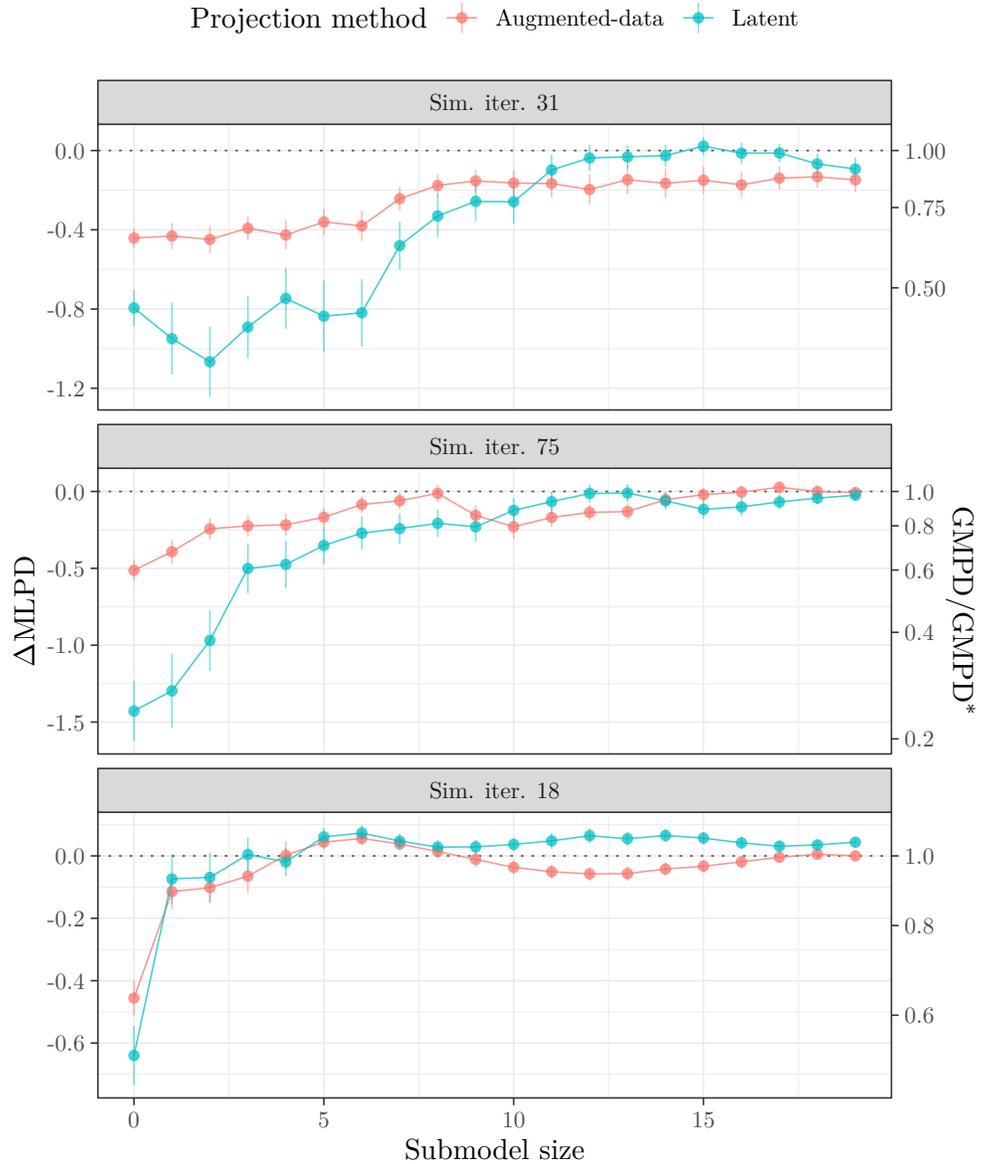}
  \end{adjustbox}
  \caption[Relative predictive performance in extreme simulation iterations]{
  Relative predictive performance at increasing submodel sizes for the augmented-data projection (light red) and the latent projection (turquoise) in the three simulation iterations (represented by panels) with the largest values for $\mathrm{MLPD}_{\mathrm{lat}} - \mathrm{MLPD}_{\mathrm{aug}}$.
  The right y-axis is simply the $\exp(\cdot)$ scale of the left y-axis.
  The vertical uncertainty bars indicate $\mathrm{SE}(\Delta\mathrm{MLPD})$
  }
  \label{fig:indiv-maxdiff}
\end{figure}

Figure~\ref{fig:indiv-maxdiff} is a restriction of \figauglat\ to the same three simulation iterations, but with a slightly different arrangement:
In Figure~\ref{fig:indiv-maxdiff}, the lines correspond to the two projection methods and the three simulation iterations are represented by panels.

What is interesting in Figures~\ref{fig:diff-maxdiff} and~\ref{fig:indiv-maxdiff} is that two of the three selected iterations (\(31\) and \(75\)) are among those where the latent projection performs extraordinarily badly at \emph{small} submodel sizes.
Thus, a bad performance of the latent projection at small submodel sizes does not necessarily imply a bad performance overall.
However, such a catch-up of the latent projection does not help if the augmented-data projection leads to a predictive performance close to the reference model at a submodel size smaller than the size where the catch-up takes place.
This is the case in iteration \(75\), but not in iteration \(31\).
In iteration~\(31\), the augmented-data projection does not manage to reach a predictive performance close to the reference model (at least not up to the maximum submodel size of~\(19\) here implied by the default of argument \texttt{nterms\_max} of \texttt{projpred::varsel()}), leaving a gap in predictive performance that the latent projection does not exhibit.

Iteration \(18\) does not exhibit a pronounced catch-up of the latent projection, but a striking spread of the two curves at larger submodel sizes.
Apparently, the augmented-data projection overfits after having attained the maximum predictive performance at size \(6\).
The latent projection overfits after this point as well, but the dip in predictive performance is shorter and not that deep.
For a judgment of the consequences of this advantage of the latent projection, it is again important to consider the submodel size where a sufficient predictive performance is reached (i.e., the submodel size which would typically be selected by the user):
Here, the major advantage in predictive performance occurs \emph{after} the point of sufficient predictive performance, and so it is not that relevant (unlike the advantage from iteration \(31\)).
However, there is also a minor advantage in predictive performance at submodel sizes \(1\) to \(3\), which is indeed relevant because it takes place \emph{before} the point of sufficient predictive performance of the augmented-data projection and even causes the \texttt{projpred::suggest\_size()} heuristic to suggest a submodel size that is smaller by three predictor terms.

As a side-effect, Figure~\ref{fig:indiv-maxdiff} shows that for larger submodel sizes in iteration~\(31\), the latent-projection SEs are smaller than their counterparts based on the augmented-data projection.
This is a rather rare case (see \figdiffse).

\section{Absolute-scale predictive performance}\label{resabs}

\figauglat\ allowed us to compare the predictive performance relative to the reference model between both projection methods.
For example, under the augmented-data projection, the submodel GMPD is always at least as large as \(50\,\%\) of the reference model GMPD whereas under the latent projection, there are also several submodel GMPDs between ca.\ \(25\,\%\) and \(50\,\%\) of the reference model GMPD.

The aim of this section is now to investigate whether the discrepancies between augmented-data and latent projection are also relevant on \emph{absolute} scale (i.e., not relative to the reference model).

Unfortunately, \figauglat\ cannot be modified easily to show the absolute scale of MLPD and GMPD.
The reason is that the reference model performance varies from simulation iteration to simulation iteration so that in a plot where the lines from all simulation iterations are combined, there would not be a \emph{single} dashed horizontal line for the reference model, but \(R = 100\) ones.
Thus, the only remedy is to inspect the results on absolute scale separately for a few simulation iterations.

To select a few iterations, we consider the difference \(\mathrm{GMPD}_{\mathrm{lat}} - \mathrm{GMPD}_{\mathrm{aug}}\) at size \(G_{\mathrm{min}} = \min(G_{\mathrm{aug}}, G_{\mathrm{lat}})\) (in the same fashion as for \figdiffsgg) and choose those iterations where this suggested-size GMPD difference is either extremely small or extremely large (taking three iterations from both extremes).

Tables~\ref{tab:indiv-mindiffexp} and~\ref{tab:indiv-maxdiffexp} show the corresponding results at size \(G_{\mathrm{min}}\).
From Table~\ref{tab:indiv-mindiffexp}, we can infer that the augmented-data projection achieves an additive suggested-size GMPD improvement (compared to the latent projection) of up to \(6.9\,\%\), with the three largest of these improvements all being between \(6\,\%\) and \(7\,\%\).
Table~\ref{tab:indiv-maxdiffexp} shows that the latent projection achieves an additive suggested-size GMPD improvement of up to \(6.3\,\%\) (similar to the maximum improvement achieved by the augmented-data projection), but the second and third largest improvements are considerably smaller than \(6\,\%\).
In general, we would consider an additive GMPD improvement between \(6\,\%\) and \(7\,\%\) as relevant, remembering that a geometric mean gives the value that could be assigned to all factors of a product (here the joint predictive probability) to arrive at the same value of the product as when taking the original factors.

\begin{table}[!tb]
\centering
\caption[Predictive performance at the minimum of the two suggested sizes for individual simulation iterations]{Predictive performance at size $G_{\mathrm{min}} = \min(G_{\mathrm{aug}}, G_{\mathrm{lat}})$ for the three simulation iterations with the smallest values for $\mathrm{GMPD}_{\mathrm{lat}} - \mathrm{GMPD}_{\mathrm{aug}}$ at size $G_{\mathrm{min}}$. Rows are sorted from most extreme (top) to least extreme (bottom)}
\label{tab:indiv-mindiffexp}
\begin{tabular}{rrrr}
  \toprule
Sim. iter. & $\mathrm{GMPD}_{\mathrm{aug}}$ & $\mathrm{GMPD}_{\mathrm{lat}}$ & $\mathrm{GMPD}_{\mathrm{lat}} - \mathrm{GMPD}_{\mathrm{aug}}$ \\
  \midrule
96 & 0.38 & 0.31 & -0.069 \\
  75 & 0.37 & 0.30 & -0.065 \\
  69 & 0.34 & 0.28 & -0.062 \\
   \bottomrule
\end{tabular}
\end{table}

\begin{table}[!tb]
\centering
\caption[Predictive performance at the minimum of the two suggested sizes for individual simulation iterations]{Predictive performance at size $G_{\mathrm{min}} = \min(G_{\mathrm{aug}}, G_{\mathrm{lat}})$ for the three simulation iterations with the largest values for $\mathrm{GMPD}_{\mathrm{lat}} - \mathrm{GMPD}_{\mathrm{aug}}$ at size $G_{\mathrm{min}}$. Rows are sorted from most extreme (top) to least extreme (bottom)}
\label{tab:indiv-maxdiffexp}
\begin{tabular}{rrrr}
  \toprule
Sim. iter. & $\mathrm{GMPD}_{\mathrm{aug}}$ & $\mathrm{GMPD}_{\mathrm{lat}}$ & $\mathrm{GMPD}_{\mathrm{lat}} - \mathrm{GMPD}_{\mathrm{aug}}$ \\
  \midrule
31 & 0.36 & 0.43 & 0.063 \\
  41 & 0.44 & 0.46 & 0.021 \\
  18 & 0.35 & 0.36 & 0.014 \\
   \bottomrule
\end{tabular}
\end{table}

\begin{figure}[!tb]
  \centering
  \begin{adjustbox}{max width=\textwidth}
  \input{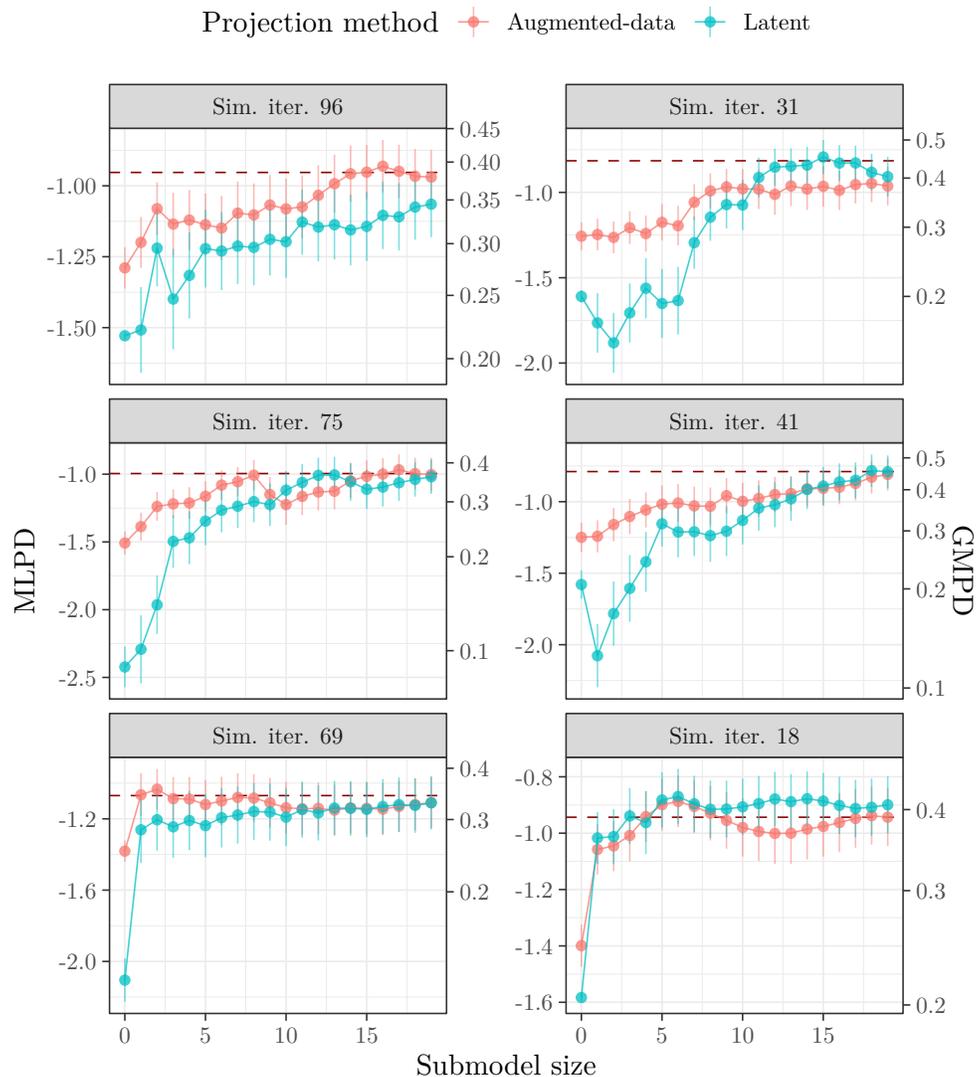}
  \end{adjustbox}
  \caption[Predictive performance for individual simulation iterations]{
  Predictive performance at increasing submodel sizes for the augmented-data projection (light red) and the latent projection (turquoise) in those simulation iterations (represented by panels) where the ``suggested-size GMPD difference'' defined in the text is either extremely small (left three panels) or extremely large (right three panels).
  Both columns are sorted from most extreme (top) to least extreme (bottom).
  The right y-axis is simply the $\exp(\cdot)$ scale of the left y-axis.
  The vertical uncertainty bars indicate $\mathrm{SE}(\mathrm{MLPD})$.
  The dashed horizontal lines indicate the predictive performance of the reference model
  }
  \label{fig:indiv-diffexp}
\end{figure}

Figure~\ref{fig:indiv-diffexp} visualizes the absolute-scale predictive performance at all submodel sizes from the forward search (not only \(G_{\mathrm{min}}\)) for all of these most extreme simulation iterations.
That visualization confirms the conclusions from Tables~\ref{tab:indiv-mindiffexp} and~\ref{tab:indiv-maxdiffexp}:
The improvements of the augmented-data projection are persistent across all three iterations from the left column, whereas the latent projection leads to a clear advantage only in the most extreme iteration (the uppermost one) from the right column.
This is iteration~\(31\) that was already discussed in Appendix~\ref{maxdiff}.
Iteration~\(18\) (the lowermost one from the right column) was already discussed in Appendix~\ref{maxdiff} as well.
Interestingly, iteration~\(75\) (the middle one from the left column) comes with the second largest additive suggested-size GMPD improvement of the \emph{augmented-data} projection, but was also discussed in Appendix~\ref{maxdiff}, meaning that it also comes with one of the largest MLPD improvements of the \emph{latent} projection, but only when considering all submodel sizes.
This is possible due to the crossing of the two curves in iteration~\(75\) (Figure~\ref{fig:indiv-diffexp}), with the augmented-data projection achieving a predictive performance close to the reference model earlier than the latent projection.
An explanation for the crossing of the two curves might be that the inclusion of at least one predictor (probably two predictors, see Figure~\ref{fig:indiv-diffexp}) causes the projection to overfit, and that the two projection methods include this predictor at different submodel sizes.

We may conclude that even at the preferable suggested size \(G_{\mathrm{min}}\), the discrepancy between augmented-data and latent projection can be relevant on absolute scale, although not huge.
Of course, the six simulation iterations were selected by cherry-picking extreme ones, but this was necessary to investigate how large the absolute-scale discrepancy at the preferable suggested size can get.
To avoid a false impression, we repeat that most of the time, the predictive performance (also on absolute scale) is similar between the two projection methods (see \figdiff).

\end{appendices}

\bibliographystyle{abbrvnat}
\bibliography{references.bib}

\end{document}